\documentclass{aa}

\usepackage{graphicx}
\usepackage{txfonts}
\usepackage{multirow}

\newcommand{\dd}{\mathrm{d}}

\usepackage[colorlinks=true,citecolor=blue,linkcolor=green]{hyperref}

\begin{document}

   \title{Improving constraints on primordial non-Gaussianity from Quaia with a new cosmological observable: Angular redshift fluctuations}

   \author{J.~R.~Bermejo-Climent\inst{1,2,3,4,}\thanks{\email{jose.bermejo@csfk.org}}
   \and 
   C. Hernández-Monteagudo \inst{1,2}
          \and
    A. Crespo-Pérez \inst{1,2}
\and
J. Martin Camalich \inst{1,2}
\and 
D. Alonso \inst{5}
\and
G. Fabbian \inst{6,7}
\and
K. Storey-Fisher \inst{8}
          }

   \institute{Instituto de Astrof\'{\i}sica de Canarias, C/ V\'{\i}a L\'{a}ctea, s/n, E-38205 La Laguna, Tenerife, Spain 
\and Departamento de Astrof\'{\i}sica, Universidad de La Laguna, Avenida Francisco S\'{a}nchez, s/n, E-38205 La Laguna, Tenerife, Spain 
\and MTA–CSFK Lendület “Momentum” Large-Scale Structure (LSS) Research Group, Konkoly Thege Miklós út 15-17, H-1121 Budapest, Hungary
\and Konkoly Observatory, HUN-REN Research Centre for Astronomy and Earth Sciences, H-1121 Budapest, Hungary 
\and Department of Physics, University of Oxford, Denys Wilkinson Building, Keble Road, Oxford OX1 3RH, United Kingdom
\and Université Paris-Saclay, CNRS, Institut d’Astrophysique Spatiale, 91405, Orsay, France
\and Kavli Institute for Cosmology Cambridge, Madingley Road, Cambridge CB3 0HA, UK
\and Kavli Institute for Particle Astrophysics and Cosmology, Stanford University, 452 Lomita
Mall, Stanford, CA 94305, USA}

\date{Received XXX; accepted YYY}
\abstract
{Angular redshift fluctuations (ARFs) are a new cosmological observable recently proposed in the literature. It measures the 2D angular deviations of the average redshift of a given matter tracer under an input redshift shell. Since it depends on galaxy bias, it can be used to constrain primordial non-Gaussianity through the scale-dependent bias effect.}
{We analyzed a sample of quasars built on Gaia satellite and unWISE data, Quaia to measure the local non-Gaussianity parameter $f_{\rm NL}$. This sample is particularly suitable for measuring $f_{\rm NL}$ due to its large volume coverage.}
{We measured the ARF power spectra from the Quaia catalog and combined their information with the 2D (projected) galaxy density and their cross-correlation with the $Planck$ PR4 cosmic microwave background lensing maps to jointly constrain $f_{\rm NL}$.}
{Assuming the universality relation, we measure $f_{\rm NL} = -3 \pm 14$ at the 68\% confidence level by combining Quaia quasar angular density and ARFs with their CMB lensing cross-correlations. Neglecting the ARF - CMB lensing cross-correlation leads to a significant improvement in the model's goodness-of-fit and yields comparable constraints, $f_{\rm NL} = -5^{+16}_{-15}$. This result is the second tightest constraint on $f_{\rm NL}$ using LSS two-point statistics to date and the best measurement achieved using two-point projected summary statistics, improving the previous measurement from Quaia by up to $\sim$25\%. Our results support the inclusion of ARFs as an additional cosmological observable in future 2D analyses of upcoming datasets from large surveys.}
{}

    \titlerunning{Improving primordial non-Gaussianity constraints from Quaia with ARF}
   \maketitle

\section{Introduction}
\label{sec:intro}
Cosmic inflation was proposed in the early 1980s \citep{Guth:1980zm,STAROBINSKY198099} as a theory for the very early universe. The inflation framework was initially formulated to solve problems associated with the Big Bang scenario such as the horizon, flatness, and magnetic monopole problems; however, inflation is also capable of explaining the formation of primordial density perturbations \citep{STAROBINSKY1982175,Guth:1985ya,Baarden}. Inflation is defined as a phase of the exponential expansion of the Universe, driven by a scalar field $\phi$. Several models of inflation have been proposed in the literature (see,  e.g., \citealt{Langlois_2010,Vazquez_Gonzalez_2020} for reviews). In these models, the choice of the potential $V(\phi)$ defines both the inflationary scenario and its predictions. The simplest inflationary models predict Gaussian initial conditions, although alternative inflationary models predict different levels of non-Gaussianity in the primordial density perturbations \citep{Chen_2010,10.1093/ptep/ptu060}. A particularly simple configuration 
of non-Gaussianity has been characterized in the literature with the $f_{\rm NL}$ non-Gaussianity parameter, where a detection of $f_{\rm NL} \neq 0$ is a signature of non-Gaussian initial conditions. In practice, the $f_{\rm NL}$ parameter describes the amplitude of a quadratic non-Gaussian term in the expression for the primordial potential.

The tightest constraint on $f_{\rm NL}$ is currently achieved with measurements from the cosmic microwave background (CMB) bispectrum. Using $Planck$ 2018 data, a value $f_{\rm NL} = -0.9 \pm 5.1$ at the 68\% confidence level was reported in \citet{planckcollaboration2019planck}.  However, \citet{PhysRevD.77.123514} first noticed that local non-Gaussian initial conditions produce a characteristic scale-dependent signature in the galaxy bias, following a $1/k^2$ scale-dependence in the ratio of the total matter to the observed galaxy density. In the last decade, many studies have measured the large-scale structure using the scale-dependent bias effect (\citealt{Ross_2012,chm_2014,Castorina_2019,muller,cabass,damico2023limits}, among others). In the past few years, LSS measurements accounting for the scale-dependent galaxy bias have been achieved using eBOSS DR16 quasars (QSOs). Measurements $f_{\rm NL} = -12 \pm 21$ in \citet{muller} and $-4 < f_{\rm NL} < 27$ in  \citet{Cagliari_2024} represent the limits obtained at the 68\% confidence level using different methodologies. This constraint was improved to $f_{\rm NL} = -3.6^{+9.1}_{-9.0}$ using the 3D power spectrum of DESI DR1 galaxies and QSOs \citep{Chaussidon_2025}. More recently, \citet{chudaykin2025reanalyzingdesidr13} claimed that the DESI DR1 constraint can be improved to $\sigma(f_{\rm NL}) \gtrsim 4$ by adding the LSS bispectrum. Measuring $f_{\rm NL}$ from LSS is often challenging because it requires a very accurate control of the observational systematics that affect the largest-scale clustering signal, where the scale-dependent bias effect arises due to $f_{\rm NL}$ (see, e.g., \citealt{Rezaie_2021,Rezaie_2024} for detailed analyses of observational systematics impacts on $f_{\rm NL}$ measurements from eBOSS and DESI data). 

The most common observable used in the literature for measuring $f_{\rm NL}$ from LSS is the 3D galaxy power spectrum in Fourier space. However, in many cases a 2D analysis in the harmonic space can be more convenient for many reasons. In particular, a 2D analysis in tomographic redshift bins can be equivalent to a 3D analysis under certain conditions (see, e.g., \citealt{Asorey_2012,Camera_2018}); it does not require us to assume any fiducial cosmology or neglect the redshift evolution inside a given redshift bin. Furthermore, possibility cross-correlations with other observables such as CMB lensing as well as the redshift uncertainties in photometric surveys make the 2D power spectrum a standard tool for the analysis of photometric galaxy counts. In this context, many recent studies have performed measurements of $f_{\rm NL}$ using 2D angular power spectra of galaxies (e.g., \citealt{Rezaie_2024}) and their cross-correlation with CMB lensing, particularly with $Planck$ PR4 CMB lensing cross-correlated with the LRG \citep{bermejo24} and the QSO \citep{Krolewski_2024,Chiarenza_26} from DESI data.

In addition, several studies in the literature \citep{Hern_ndez_Monteagudo_2020,10.1093/mnrasl/slab021,Lima_Hern_ndez_2022,Matthewson_2022} have proposed a new 2D observable called ``angular redshift fluctuations" (hereafter ARFs) that can be combined with galaxy density and CMB information to constrain cosmological parameters. Angular redshift fluctuations are a new observable that introduce redshift as a 2D field and measure the 
deviations of the mean redshift of the galaxies in each direction of the sky with respect to the average redshift of the matter tracer sample under any given redshift shell. It has been shown that this observable can be useful for constraining cosmological parameters in combination with standard angular density fluctuations from theoretical forecasts \citep{Legrand_2021} and also with real data applications for measuring structure growth and gravity \citep{10.1093/mnrasl/slab021,chm_on_JPLUS_DR3}. 

As theoretically discussed in \citet{Hern_ndez_Monteagudo_2020}, ARFs are sensitive to galaxy bias and, hence, this new observable can be used to measure local primordial non-Gaussianity (PNG) $f_{\rm NL}$ thanks to the scale-dependent bias effect introduced in \citet{PhysRevD.77.123514}. In this work, we aim to explore the sensitivity of ARFs to $f_{\rm NL}$ using real cosmological datasets. For this purpose, we used the Quaia catalog \citep{Storey_Fisher_2024}, a QSO sample extracted from the Gaia DR3 observations and unWISE infrared data \citep{unwise}. We also added CMB lensing information from the $Planck$ PR4 data release to perform a cross-correlation analysis. The Quaia catalog has been already used to constrain structure growth \citep{Alonso_2023}, to measure the power spectrum turnover scale \citep{Alonso_2025} and the primordial non-Gaussianity parameter \citep{fabbian2025} cross-correlated with $Planck$ CMB lensing. Despite the low number density of the QSO, it is a full sky catalog covering the largest volume covered by any such catalog. This makes it a suitable dataset for primordial non-Gaussianity measurements given that the $f_{\rm NL}$ arises at the largest scales. Our aim in this work bifold: we seek to extend the analysis by \cite{fabbian2025} with the inclusion of ARFs as a cosmological observable and to quantify the potential improvement on the $f_{\rm NL}$ uncertainty achieved by adding ARFs to future analyses of upcoming LSS surveys such as DESI, Euclid, and LSST.

This paper is organized as follows. In Sect.~\ref{sec:theory} we describe the formalism related to the scale-dependent bias induced by a primordial local non-Gaussianity and the projected 2D observables included in our analysis. In Sect.~\ref{sec:datasets} we describe the datasets used for this measurement. In Sect.~\ref{sec:pipeline} we detail the methodology for our analysis pipeline. In Sect.~\ref{sec:results} we present our results and robustness tests on the $f_{\rm NL}$ measurement, and in Sect.~\ref{sec:conclusions} we summarize our conclusions.

\section{Theory}
\label{sec:theory}
In this section we first describe the physical model that gives rise to a scale-dependent galaxy bias due to PNG. Then, we outline the cosmological observables in the 2D harmonic space relevant to our analysis.

\subsection{Primordial non-Gaussianity and scale-dependent bias}
\label{sec:png}
If we assume a type of non-Gaussianity that depends only on the local value of the potential, the parameterization of the primordial potential can be written as follows \citep{PhysRevD.63.063002}:
\begin{equation}
    \Phi = \phi + f_{\rm NL} ( \phi^2 - \langle \phi \rangle ^2) \,,
\end{equation}
where $f_{\rm NL}$ is the parameter that describes the amplitude of the non-Gaussian quadratic term and $\phi$ is a random Gaussian field. We studied $f_{\rm NL}$ through its impact on the scale-dependent galaxy bias, as first introduced in \citet{PhysRevD.77.123514,Matarrese_2008}. If we assume the so-called universality relation for the halo mass function \citep{Slosar_2008}, the added contribution to the galaxy bias is given by
\begin{equation}
\label{deltab}
\Delta b(k,z) = 2(b_g-p) f_{\rm NL}\frac{\delta_{\rm crit}}{\alpha(k)} \,,
\end{equation}
where $\delta_{\rm crit}$ = 1.686 is the threshold overdensity for spherical collapse, $b_g$ is the $z$-dependent galaxy bias, $p$ is a parameter that characterizes the response of QSOs to PNG and $\alpha(k)$ is the relation between potential and density field, where $\delta(k,z) = \alpha(k,z) \Phi(k)$. The value of $\alpha (k,z)$ is given by
\begin{equation}
    \alpha(k,z) = \frac{2 k^2 T(k) D(z)}{3 \Omega_{\rm m}} \frac{c^2}{H_0^2} \frac{g(0)}{g (\infty)} \,,
\end{equation}
where $T(k)$ is the transfer function, $D(z)$ is the growth factor (normalized to be 1 at $z=0$), $\Omega_{\rm m}$ the matter density, and $g \equiv D/a$, such that the factor $g(\infty)/g(0) \simeq 1.3$ accounts for the different normalizations of $D(z)$ in the CMB and LSS literature. This definition of $f_{\rm NL}$ is, therefore, the so-called "CMB convention." Note that other authors \citep{Carbone_2008,Afshordi_2008,Grossi_2009} have adopted "LSS convention," where the $g(\infty)/g(0)$ factor is absorbed into the definition of  $f_{\rm NL}$, such that $f_{\rm NL}^{\rm LSS} \simeq 1.3 f_{\rm NL}^{\rm CMB}$. Throughout this work, we fixed $p = 1$ as a baseline; however, many studies based on dark-matter-only simulations (e.g., \citealt{adame}) have found significant deviations from $p=1$. In this context, some studies \citep{Barreira_2020,Barreira_2022} have stressed that we can only constrain the product $b_\phi f_{\rm NL}$ through the scale-dependent bias effect -- where  $b_\phi$ is a parameter typically defined as $b_\phi = 2 \delta_{\rm crit} (b_g - p)$ -- in order to account for the uncertainties on $p$. 

\subsection{Cosmological observables}

Our cosmological observables are the angular auto- and cross-correlation power spectra of three fields: CMB lensing, the galaxy number counts, and ARFs. The angular power spectrum can be calculated as
\begin{equation} \label{eqn:APS}
    C_\ell^{XY} = 4\pi \int \frac{\dd k}{k} {\cal P}(k) I^X_{\ell}(k) I^Y_{\ell}(k),
\end{equation}
where ${\cal P}(k) \equiv k^3 P(k) / (2\pi^2)$ is the dimensionless primordial power spectrum
and $I^X_{\ell}(k)$ is the kernel of the field $X$.

Cosmic microwave background lensing quantities can be defined from the lensing potential
\begin{equation}
    \phi\left(\hat{\bf n},\chi\right) = \frac{2}{c^2} \int_0^\chi \dd \chi' \frac{\chi-\chi'}{\chi\chi'}
    \Phi\left(\chi'\hat{\bf n},\chi'\right),
\end{equation}
where $\Phi\left(\hat{\bf n},\chi\right)$ is the gravitational potential,
and the comoving distance given by
\begin{equation}
    \chi(z) = \int_0^z \frac{c\, \dd z'}{H(z')}\,. 
\end{equation}

By expanding the gravitational potential in Fourier space and using the plane-wave expansion, we define the CMB lensing potential kernel as
\begin{equation}
 \label{eqn:kernel_CMB}
    I^{\phi_{\rm CMB}}_\ell (k) = 2\left(\frac{3\Omega_mH_0^2}{2k^2c^2}\right) \int \frac{\dd \chi}{(2\pi)^{3/2}} \frac{\chi_*-\chi}{\chi_*\chi} \frac{1}{a(\chi)}
    j_\ell\left(k\chi\right) \delta\left(k,\chi\right) \,,
\end{equation}
where $\Omega_m$ is the present-day matter density, $H_0$ is the Hubble constant, $\delta(k,\chi)$ 
is the synchronous gauge linear matter density perturbation, $\chi_*$ is the comoving distance at the surface of last scattering, and $j_\ell$ the spherical Bessel functions. 

Finally, the convergence $\kappa = \nabla^2 \phi/2$ is expanded in spherical harmonics as
\begin{equation}
    \kappa\left({\bf \hat{n}}\right) = -\frac{1}{2} \sum_{\ell, m} \ell(\ell+1)\phi_{\ell m}Y_\ell^{m}\left({\bf \hat{n}}\right).
\end{equation}
We relate the two kernels by
\begin{equation} \label{eqn:kernel_kappa}
    I_\ell^{\kappa}(k) = \frac{\ell(\ell+1)}{2} I_\ell^{\phi}(k) \,.
\end{equation}

The 2D integrated window function for the galaxy number counts is given by
\begin{equation} \label{eqn:kernel_counts}
    I^g_\ell(k) = \int \frac{\dd \chi}{(2\pi)^{3/2}} W(\chi) \Delta_{\ell}(k,\chi), 
\end{equation}
where $\Delta_{\ell}(k,\chi)$ is the observed number counts and $W(\chi)$ is a window function given by the redshift distribution of galaxies. At first order, the most important contribution to  $\Delta_{\ell}(k,\chi)$ is given by the synchronous gauge source counts Fourier transformed and expanded into multipoles, 
$\Delta^s_{\ell}(k,\chi)$. In our study, we assumed that $\Delta^s_{\ell}(k,\chi)$ is related to the underlying matter density field through a scale and redshift-dependent galaxy bias $b_g$ via 
\begin{equation}
    \Delta^s_{\ell}(k,\chi) = b_g(k,\chi) \delta(k,\chi) j_\ell\left(k\chi\right),
\end{equation}
where $b_g(k,\chi)$ is given by the sum of a linear bias (not scale-dependent) plus the scale-dependent contribution given by Eq.~\ref{deltab}. We also considered nonlinear contributions to the power spectrum using \texttt{halofit} \citep{Takahashi_2012}.
We did not include corrections from general relativity to the galaxy number counts. According to \cite{Alonso_2023} the measured magnification bias $s$ of the Quaia catalog for the $G < 20.5$ magnitude limit cut is expected to be $s \sim 0.4$, and this is the value for which the lensing magnification contribution vanishes. Remaining general relativity corrections have less impact on galaxy number counts and their cross-correlations (e.g., \citealt{Bermejo_Climent_2021}). In particular, their effects are expected to be negligible for primordial non-Gaussianity measurements \citep{Alonso_2015,Guedezounme_2025}.

The ARF field is formally defined as in \cite{Hern_ndez_Monteagudo_2020}:
\begin{equation}
\label{eq:arf1}
   (\delta z)\left({\bf \hat{n}}\right) = \frac{\sum_{j \in \hat{\bf n}} (z_j -  \bar{z}) w_j}{\sum_{j \in \hat{\bf n}} w_j },
\end{equation}
where the $j$ index runs over the galaxies pointing to a given direction $\hat{\bf n}$ and $\bar{z}$ is the average redshift of galaxies under a user-defined redshift window $W$. This latter value can be arbitrary, but in many cases we adopted a Gaussian form for simplicity using an associated weight for each galaxy given by
\begin{equation}
    w_j \equiv \exp{ \left[ - \frac{(z_{\rm obs} - z_j)^2}{2 \sigma_z^2} \right] }\,,
\end{equation}
 with $z_{\rm obs}$ and $\sigma_z$ corresponding to the central redshift and the width of the Gaussian window. In this particular case, we chose to adopt a very wide Gaussian window, which is equivalent to selecting galaxies under the survey's intrinsic QSO redshift distribution. Thus, $w_j=1$ in practice, and $\bar{z}$ is the average redshift of the sample.
The definition of the ARF field in Eq.~\ref{eq:arf1} can be noisy when the number of galaxies per pixel is low or even zero. In this study, for practical reasons related to the limited number of QSO per unit of observed area, we adopted a second definition of the ARF field as in \cite{10.1093/mnrasl/slab021}:
\begin{equation}
\label{eq:arf2}
    (\delta z)\left({\bf \hat{n}}\right) = \frac{\sum_{j \in \hat{\bf n}} (z_j - \bar{z}) w_j}{\langle \sum_{j \in \hat{\bf n}} w_j \rangle_{\hat{\bf n}}} \,,
\end{equation}
where we took the ensemble average of the density over all the survey footprint in the denominator. We note that the definition in Eq.~\ref{eq:arf1} is insensitive to multiplicative systematics, while both definitions are sensitive to additive systematics, assuming no redshift dependence in the systematics under the redshift window $W$.

To calculate the ARF field kernel $I_\ell^{z}$, we followed the same formalism for the galaxy number counts but replacing the window function $W$ in Eq.~\ref{eqn:kernel_counts} with a redshift-fluctuation-weighted window $\cal{W}$,
\begin{equation}
    {\cal W} (z,\bar{z}) \equiv  W(z) (z-\bar{z}),
\end{equation}
as implemented in the modified version of the Boltzmann code {\tt CAMB} \citep{camb2} named {\tt ARFCAMB}\footnote{{\tt ARFCAMB} can be accessed at \url{https://github.com/chmATiac/ARFCAMB}} \citep{Lima_Hern_ndez_2022}. We note that although the Eq.~\ref{eqn:kernel_counts} integral is in $\chi$, the input for the {\tt ARFCAMB} code is given as a window function in redshift $z$, which the code in turn internally transforms to the comoving distance $\chi$.
\section{Datasets}
\label{sec:datasets}
In this section, we present the two main ingredients of our analysis. These are the Quaia QSO catalog, extracted from Gaia data, which contains approximately $\sim$ 1.3 million QSO, and the $Planck$ PR4 release of the CMB lensing maps.  

\subsection{The Quaia QSO catalog}
Quaia \citep{Storey_Fisher_2024} is a QSO catalog constructed from the Gaia DR3 QSO candidates sample and unWISE \citep{Schlafly_2019} infrared data. It is an all-sky catalog that covers the largest volume of any existing spectroscopic QSO sample. The full sample, with $G < 20.5$ limiting magnitude, contains 1,295,502 sources. A cleaner version exists with a $G < 20.0$ limiting magnitude and 755,850 QSOs. The spectro-photometric redshifts in the catalog were improved by training a k-nearest neighbors model on SDSS redshifts, achieving estimates on the $G < 20.0$ sample with only $\sim10\%$ catastrophic redshift errors, such that $\lvert \Delta z/(1+z)\rvert > 0.1$. In our case, we used the $G < 20.5$ sample, for which $\sim70\%$ of the sources agree to $\lvert \Delta z/(1+z)\rvert < 0.1$ and $\sim$62\% to $\lvert \Delta z/(1+z)\rvert < 0.01$.

\begin{figure}
\centering
\includegraphics[width=\columnwidth]{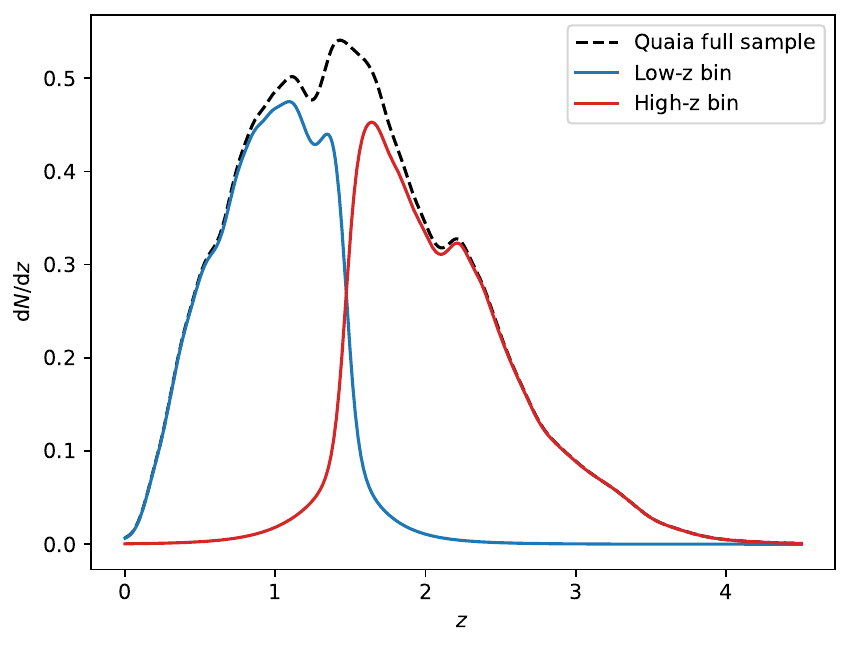}
    \caption{Normalized redshift distribution of the Quaia QSO sample analyzed in this paper. The dashed black line corresponds to the full sample, while the blue and red lines correspond to the low- and high-redshift bins, respectively.}
    \label{fig:dndz}
\end{figure}
Together with these catalogs, the following are provided: a HEALpix selection function map, which gives the probability that a source in a given pixel is included in the catalog; a random catalog, downsampled according to the selection function and containing about ten times the number of Quaia QSO; and observational systematics templates. For our analysis, we split the Quaia $G < 20.5$ sample in two redshift bins as in \cite{fabbian2025}. These two redshift bins are equally populated, splitting the sample around $z \sim 1.5$ with average redshifts $\bar{z} = 0.97, \bar{z} = 2.10$. We imposed a selection function threshold of 0.5 to exclude low completeness regions close to the galactic plane, which could introduce systematics. To account for the redshift uncertainties when splitting the sample in two redshift bins, we computed the redshift distributions as the sum at every redshift of Gaussian photo-z functions for each QSO given its observed redshift and redshift uncertainty. Accounting for the redshift uncertainties is also important in particular for the ARF theory estimations, since it decreases the measured signal. In this case we took into account the average redshift uncertainty of the Quaia $G < 20.5$ catalog, which is $\sigma_z \simeq 0.06(1+z)$ for the theoretical ARF model (see Sect.~\ref{sec:model} for more details).
We show in Fig.~\ref{fig:dndz} the redshift distribution of the two $G< 20.5$ redshift bins together with the full sample distribution.   
The catalogs, the selection functions, and all related materials are publicly available\footnote{\url{https://zenodo.org/records/10403370}}.

\subsection{$Planck$ CMB lensing}

The second ingredient in our analysis is the $Planck$ CMB lensing potential map. We used the $Planck$ PR4 reconstruction of the CMB lensing potential \citep{Carron_2022}, obtained from $Planck$ NPIPE temperature and polarization maps \citep{2020}. In particular, we used the minimum-variance estimate from these maps, after mean-field subtraction of the lensing convergence. This latest release of the CMB lensing maps improves the noise from previous $Planck$ PR3 maps \citep{1807.06209}. In particular, the large-scale noise is lower, and the mean-field is better understood thanks to the use of a larger number of simulations. The maps and mask are publicly available\footnote{\url{https://github.com/carronj/planck\_PR4\_lensing/releases/tag/Data}}. 

Note that the CMB lensing map does not include the Monte Carlo multiplicative correction applied in \cite{Carron_2022}. We computed this correction based on the simulations in \cite{fabbian2025} using mode decoupled pseudo-$C_\ell$ and applied the result as a multiplicative correction to the measured cross-correlations with the CMB lensing power spectra, $C_\ell^{\kappa g}$ and  $C_\ell^{\kappa z}$. This correction cannot be applied in a general way, since it depends on the footprint mask for each tracer involved in the analysis due to local variations of the normalization. The order of this correction is generally $\lesssim$ 5\%; however, it is greater ($\gtrsim$10\%) for the lowest multipoles, since -- for this reason -- it has a nonnegligible impact on the $f_{\rm NL}$ measurement.

\begin{figure*}
\centering
\includegraphics[width=\columnwidth]{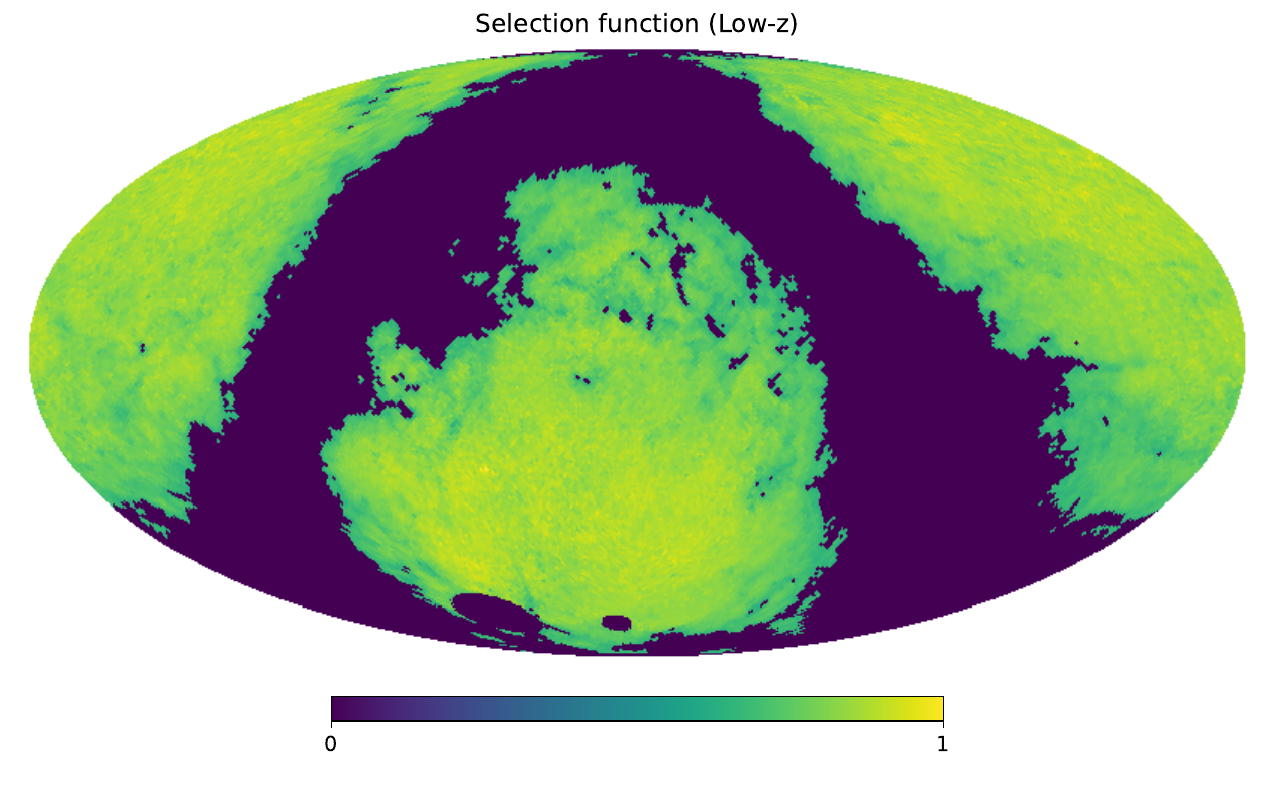}
\includegraphics[width=\columnwidth]{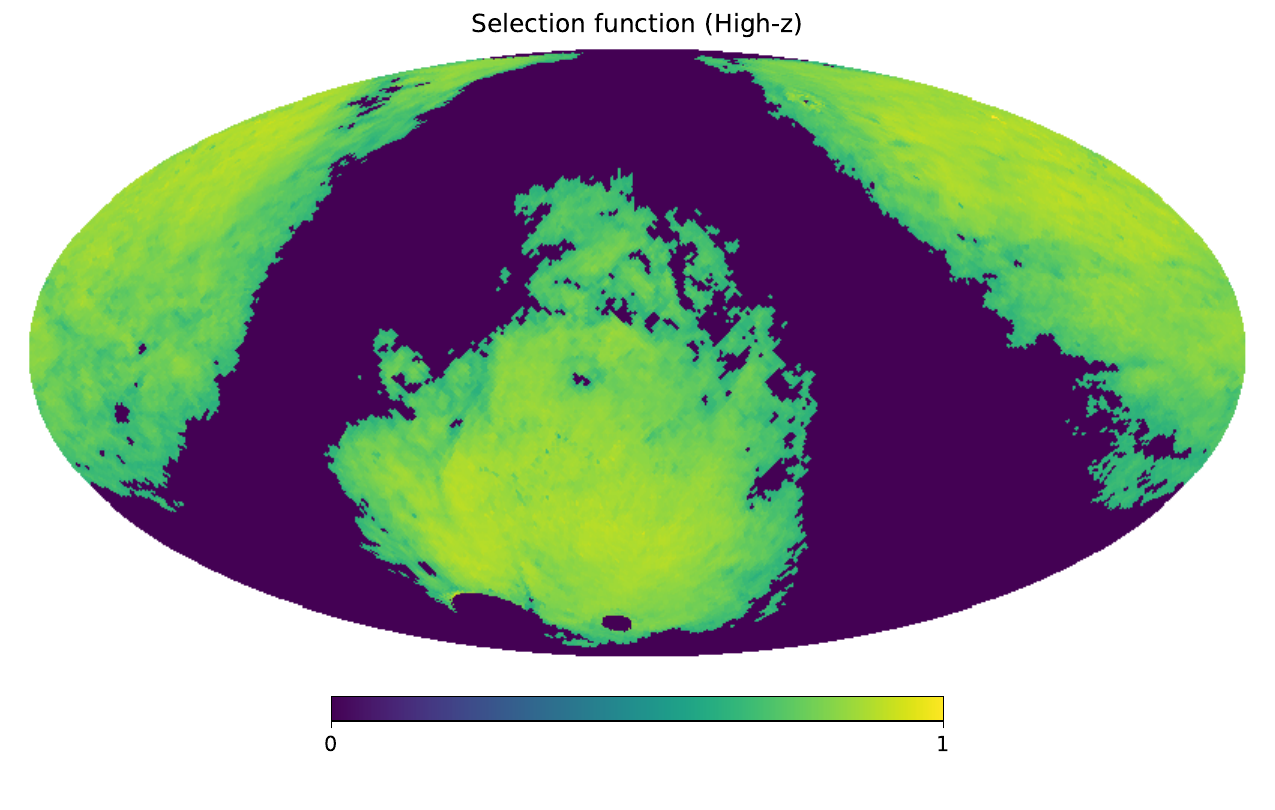}
\includegraphics[width=\columnwidth]{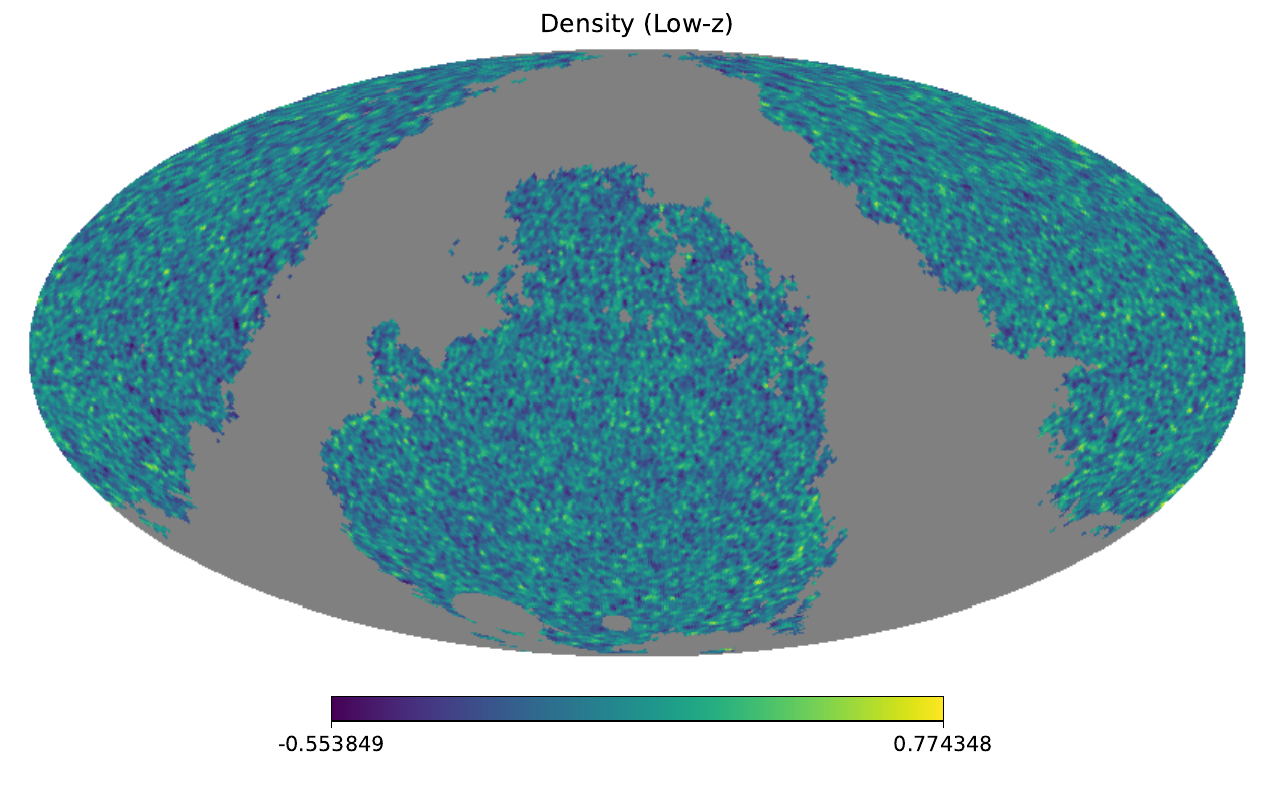}
\includegraphics[width=\columnwidth]{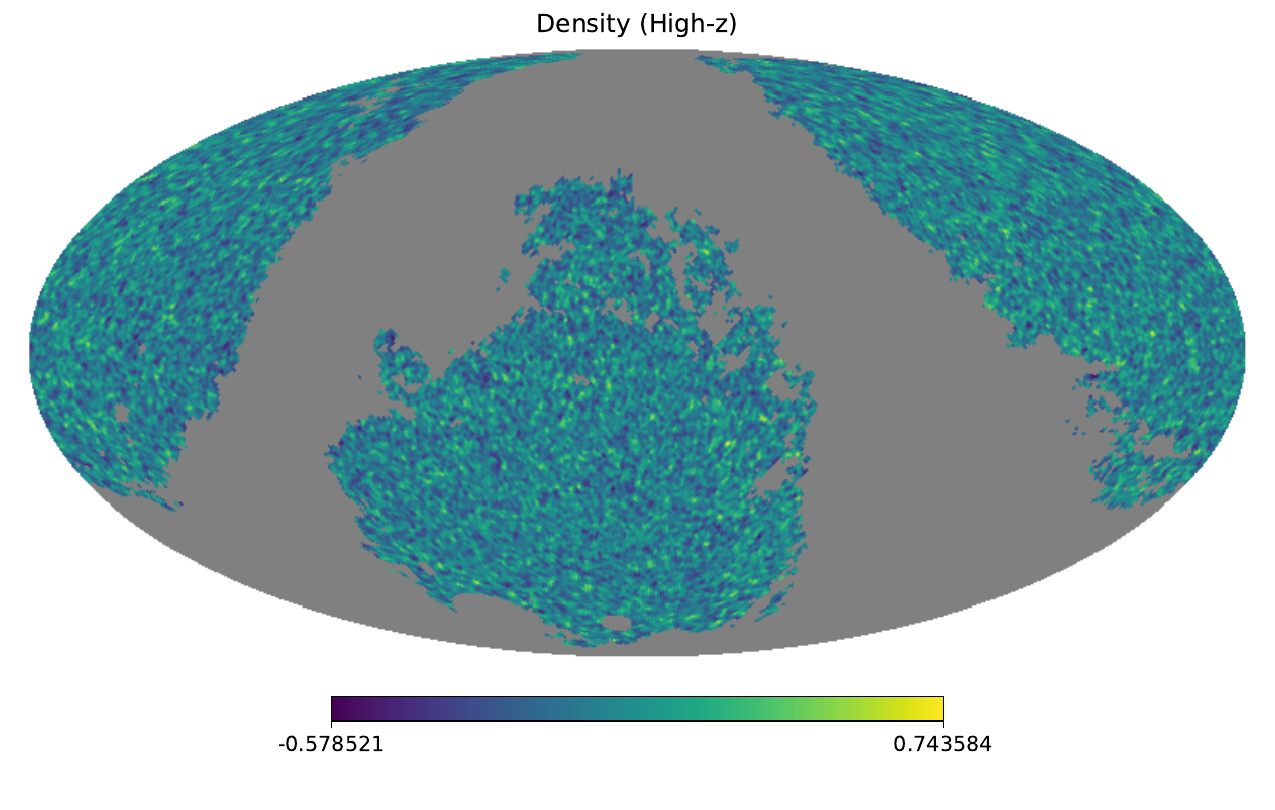}
\includegraphics[width=\columnwidth]{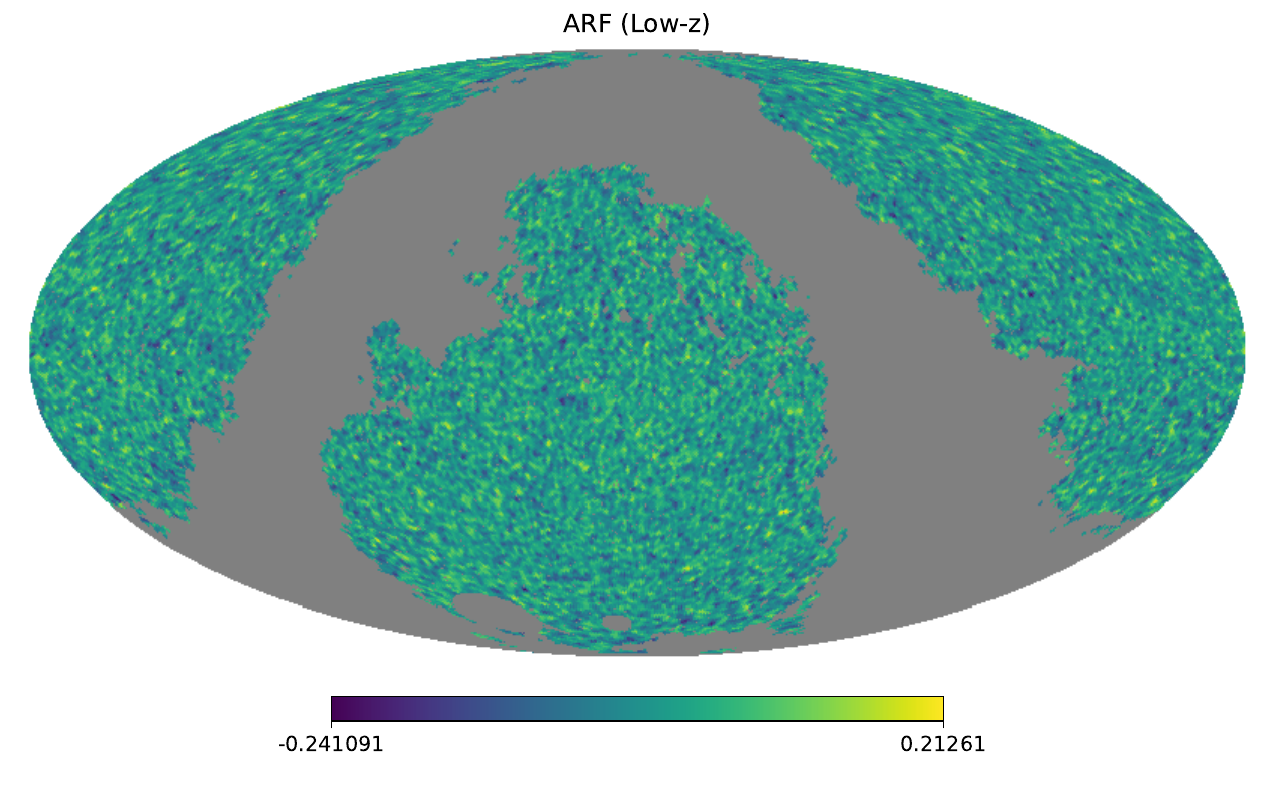}
\includegraphics[width=\columnwidth]{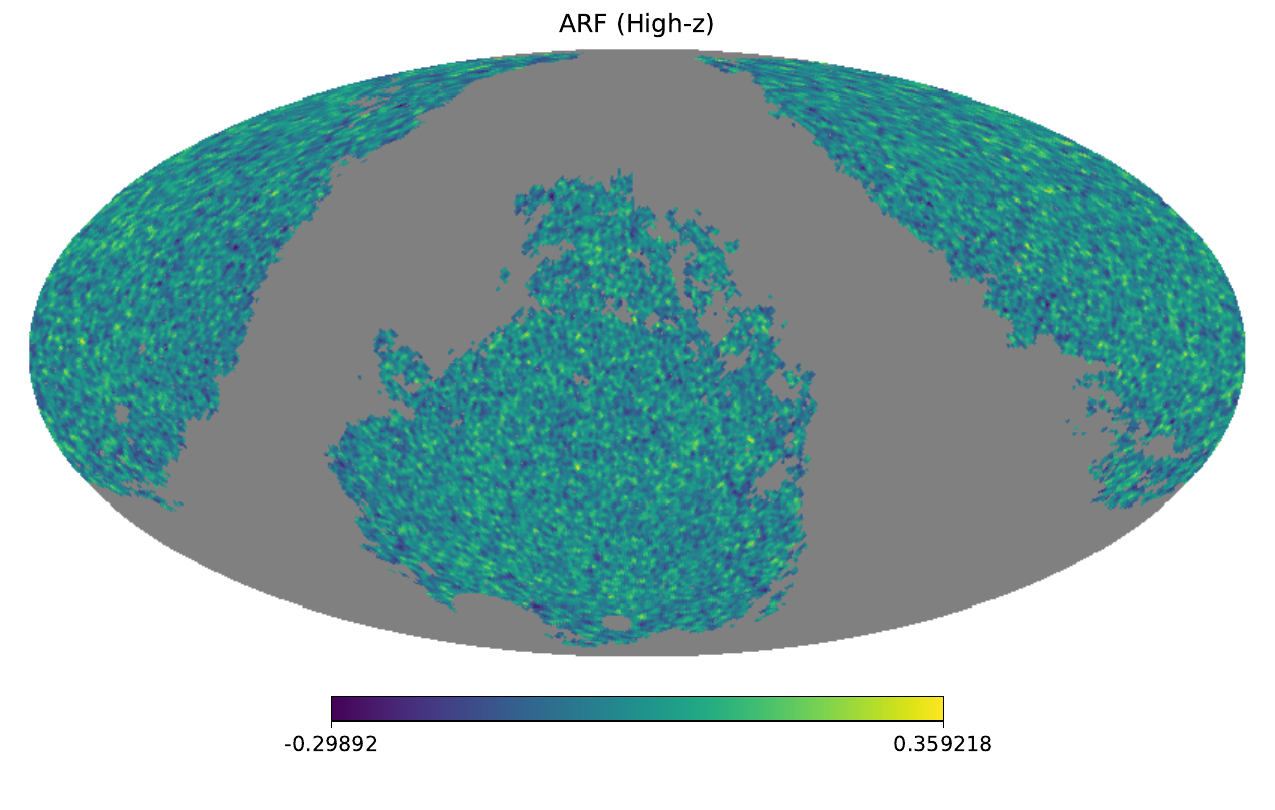}
    \caption{Upper panels: Normalized masks from the Quaia selection functions applied in our analysis after applying a 0.5 threshold for the low-z and high-z redshift bins. Middle panels: Quaia density maps for the low-z and high-z redshift bins. Lower panels: Quaia ARF maps for the low-z and high-z redshift bins. The density and ARF maps are represented with a 1 degree beam smoothing.}
    \label{fig:maps}
\end{figure*}

\section{Analysis pipeline}
\label{sec:pipeline}
In this section we describe the pipeline implemented to measure the density and ARF observables from the Quaia catalog and jointly analyze them, together with the $Planck$ CMB lensing data. The steps include creating maps and measuring the 2D angular power spectra, using a modification of the CAMB Boltzmann code  \citep{2011ascl.soft02026L} in order to generate a theoretical model, generation of correlated mock fields for computing the covariance matrix, a validation of scale cuts through systematics deprojection, and a Markov chain Monte Carlo (MCMC) likelihood code for parameter inference. 

We began by creating HEALpix \citep{Gorski_2005} maps of the density and ARF fields using the $G < 20.5$ limiting magnitude Quaia QSO catalog. For the density maps, we computed the galaxy overdensity field in each pixel as $\delta_g = \rho / \bar{\rho} - 1$, where $\rho$ is the QSO number counts after being corrected for the Quaia selection function. We imposed a threshold of 0.5 in the selection function and masked out the sky regions with lower values to avoid low-completeness areas, which could potentially add systematics to the analysis. The final footprint after the selection function cut covers $\sim 60\% $ of the sky. For the ARF maps, the redshift fluctuation field in each pixel is defined as $\delta z = \sum_i (z_i - \bar{z})/\bar{\rho}$. We created both maps using a resolution $N_{\rm side} = 128$. This choice was made to avoid empty pixels without any QSO, required for properly computing the ARF estimator. This procedure was implemented for each of the low-redshift and high-redshift bins shown in Fig.~\ref{fig:dndz}. We show the final masks and maps of the density and ARF fields for the low-$z$ and high-$z$ redshift bins Quaia sample in Fig.~\ref{fig:maps}.

\subsection{Angular power spectra}
\label{sec:aps}

To estimate the angular power spectra of the Quaia density, ARFs, their cross-correlation, as well as the cross-correlations with the $Planck$ lensing, we used the pseudo-$C_\ell$ approach implemented in the publicly available \texttt{NaMaster} code \citep{1809.09603}. The pseudo-$C_\ell$ of a pair of fields is defined as
\begin{equation}
    \Tilde{C}_\ell^{XY} = \frac{1}{2 \ell + 1} \sum_m X_{\ell m} Y_{\ell m}^{*},
\end{equation}
where $X, Y \equiv g, z, \kappa$ are the observed fields. Then, the difference between the true and measured $C_\ell$ due to the effects of the mask was accounted for using the mode-coupling matrix $M_{\ell \ell'}$:
\begin{equation}
    \langle \Tilde{C}_\ell \rangle = \sum_{\ell'} M_{\ell \ell'} C_{\ell'} \,.
\end{equation}
 In practice, $M_{\ell \ell'}$ matrix inversion is performed using the MASTER algorithm \citep{Hivon_2002}, which requires a discrete binning of the angular power spectrum. We used the implementation in the \texttt{compute\_full\_master} function in the \texttt{NaMaster} code to calculate the angular power spectra, and we binned the theory curves using the same band-power window functions. For our analysis, we set a multipole binning of $\Delta \ell = 2$ for $\ell \leq 40$, $\Delta \ell = 5$ for $40 < \ell \leq 60$, and $\Delta \ell = 30$ for $\ell > 60$. This binning scheme is also used in \cite{Alonso_2025,fabbian2025} and provides sufficiently accurate sampling of the angular power spectra at the lower multipoles where the $f_{\rm NL}$ signal arises. For the masks, we used the selection function with a 0.5 threshold for the Quaia observables and the $Planck$ PR4 mask for CMB lensing.

Note that the $Planck$ PR4 CMB lensing maps are provided at a resolution of $N_{\rm side} = 2048$, while we used Quaia maps at a resolution of $N_{\rm side} = 128$. To overcome this issue when computing the Quaia cross-correlations with CMB lensing, $C_\ell^{\kappa g}$ and $C_\ell^{\kappa z}$, we filtered out multipoles greater than $3 N_{\rm side} -1$ from the $Planck$ lensing $a_{\ell m}$ and generated a lensing map from this filtered $a_{\ell m}$ at a resolution of $N_{\rm side} = 128$. We did not use any mask apodization for the Quaia observables but rather apodized the $Planck$ mask using the \texttt{mask\_apodization} routine in \texttt{NaMaster} with the "C1" apodization type and a 0.2 degree scale. This choice was made to maintain consistency with the analysis by \citet{fabbian2025}, though its impact at our $N_{\rm side} = 128$ resolution is mostly small or negligible.

\begin{figure*}
\centering
\includegraphics[width=\textwidth]{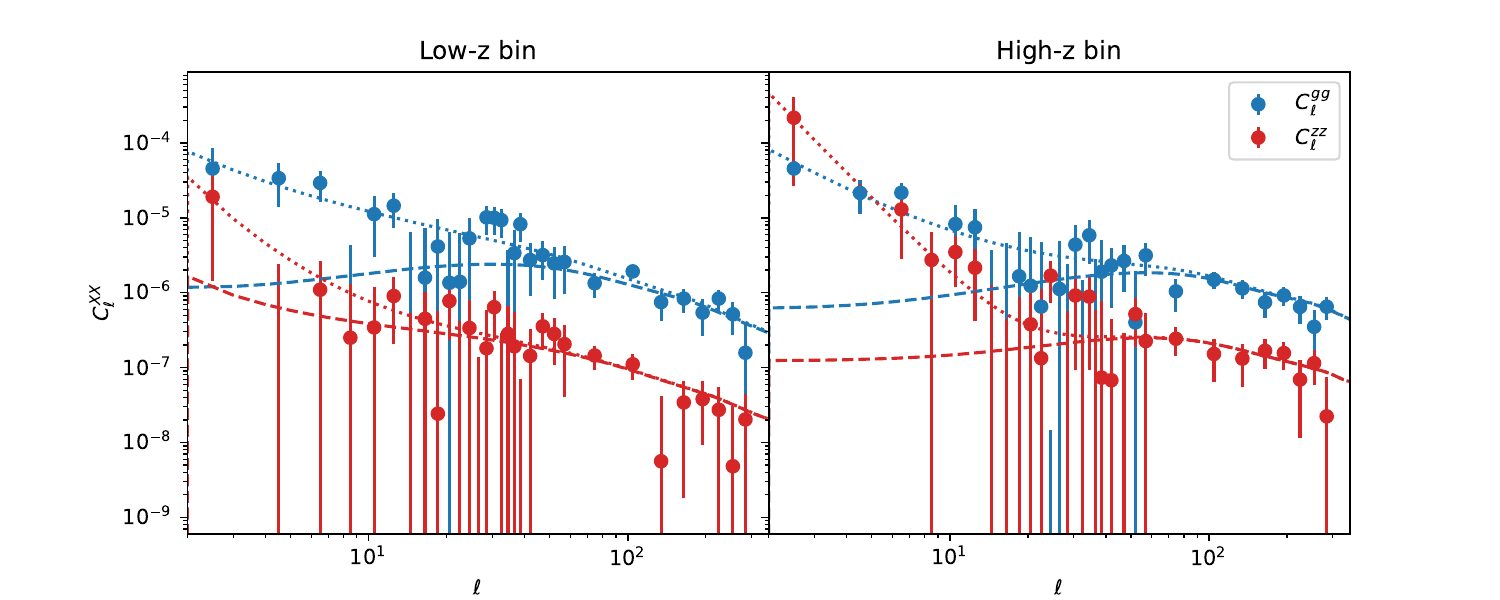}
    \caption{Measured angular power spectra of the Quaia density and ARF autocorrelations for the low-redshift (left panel) and high-redshift (right panel) bins. The dots represent the binned angular power spectra with error bars, the dashed lines represent the theoretical model with $f_{\rm NL} = 0$, and the dotted lines represent the same model after the parameterization of the measured excess of power following Eq.~\ref{eq:sys}.} 
    \label{fig:auto}
\end{figure*}

\begin{figure*}
\centering
\includegraphics[width=\textwidth]{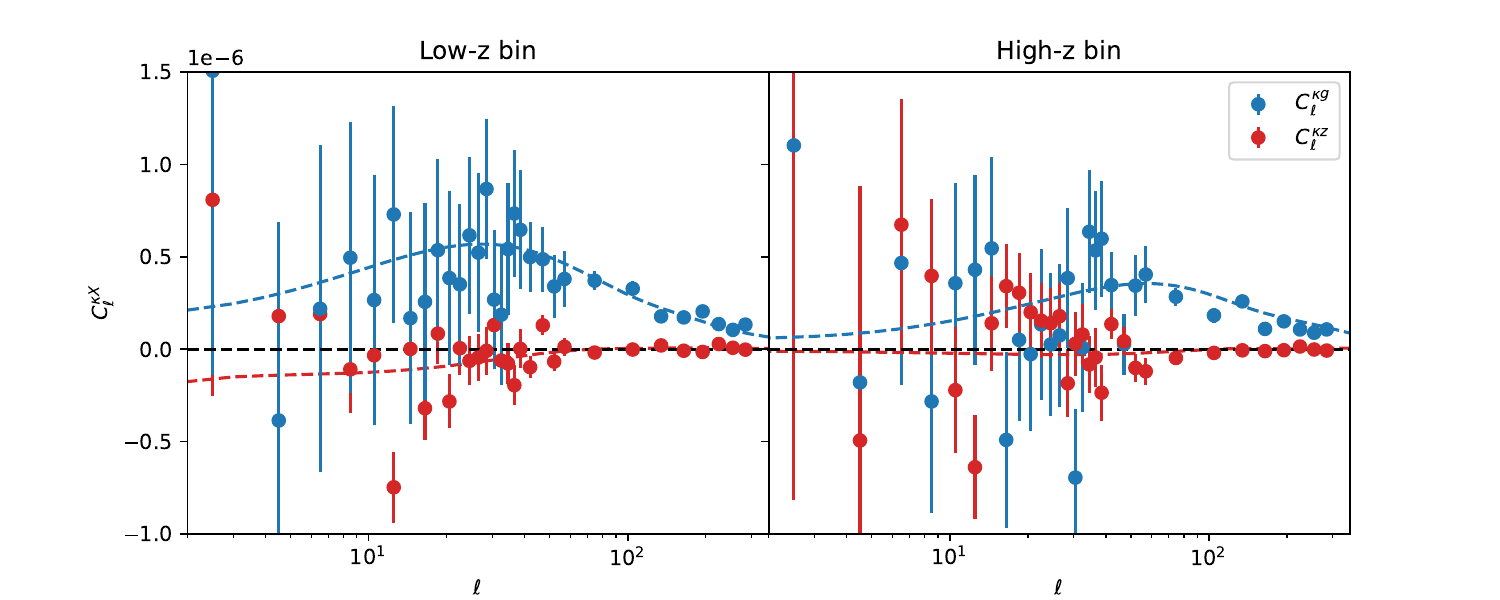}
    \caption{Measured angular power spectra of the Quaia density and ARF cross-correlations with the $Planck$ CMB lensing for the low-redshift (left panel) and high-redshift (right panel) bins. The dots represent the binned angular power spectra with error bars, and the dashed lines represent the theoretical fiducial model with $f_{\rm NL} = 0$.}
    \label{fig:cross}
\end{figure*}

Given the low QSO density of our sample, we computed and subtracted the Poisson shot noise from the autocorrelation power spectra of the density and ARFs, $C_\ell^{gg}$ and $C_\ell^{zz}$. For this, we estimated the shot noise by computing the density and ARF angular power spectra from 25 random catalogs generated according to the Quaia selection functions. We show in Fig.~\ref{fig:auto} the autocorrelation power spectra $C_\ell^{g g}$ and $C_\ell^{z z}$, as well as the CMB lensing cross-correlation power spectra in Fig.~\ref{fig:cross} ($C_\ell^{\kappa g}$ and $C_\ell^{\kappa z}$). In this paper, we do not include the cross-correlation between density and ARFs, $C_\ell^{gz}$, as an observable since it is a very weak signal that deviates relative to the theoretical model even at high multipoles. We find this indicates correlated systematics. However, we note that $C_\ell^{gz}$ was included in the calculation of the covariance matrix of our observables.

\subsection{Theoretical model}
\label{sec:model}

Our theoretical model for the angular power spectra is based on a modification of the \texttt{CAMB}  Boltzmann code 
\citep{2011ascl.soft02026L}. This modification includes the computation of ARFs as a new observable, in addition to the already existing LSS observables in the code (galaxy number counts, weak lensing, and 21 cm intensity). It also includes the possibility of introducing the width of the photometric redshift (photo-$z$) probability density distribution (PDF) under the Gaussian and Lorentzian approximations, since photo-$z$ errors considerably affect the amplitude of the ARF angular power spectra \citep{chm_on_JPLUS_DR3}. For the case of the Quaia $G < 20.5$ sample analyzed in this work, we accounted for the average redshift uncertainty of the catalog, $\sigma_z \simeq 0.06(1+z)$, in the ARF theory computations. The code is called \texttt{ARFCAMB}; it includes $f_{\rm NL}$ as an input parameter, which contributes to the scale-dependent bias. More details about the \texttt{ARFCAMB} code 
can be found in \cite{Lima_Hern_ndez_2022}. 

We generated the theoretical angular power spectra of our observables and their cross-correlations using the \texttt{ARFCAMB} code assuming a fiducial $f_{\rm NL} = 0$. The remaining cosmological parameters were fixed to the $Planck$ 2018 cosmology \cite{1807.06209}, i.e., $\Omega_{\rm b} h^2 = 0.022383$, $\Omega_{\rm c} h^2 = 0.12011$, $H_0 = 67.32$ km s$^{-1}$ Mpc$^{-1}$, and $\sigma_8 = 0.812$.
For the window function of each redshift bin, we used the d$N$/d$z$ represented in Fig.~\ref{fig:dndz}  and a fiducial evolution of the galaxy bias, as measured in \citet{Piccirilli_2024} (hereafter P24), for the Quaia sample. This evolution is parameterized through the expression
\begin{equation}
\label{eq:bzquaia}
b_{\rm QSO} (z) = b_0 / D(z)  \,,
\end{equation}
with $b_0$ = 1.26 as fiducial value. As a consistency check, we also considered the fiducial redshift evolution of the galaxy bias based in the eBOSS QSO model by \citet{Laurent_2017} (hereafter L17):
    \begin{equation}
    \label{eq:bz}
        b_{\rm QSO} (z) = 0.278 \left( (1+z)^2 - 6.565\right) + 2.393  \,.
    \end{equation}

\subsection{Mocks and covariance matrix}
\label{sec:mocks}
We generated 1,000 correlated simulations for the density, ARFs, and CMB lensing fields to compute the covariance matrices and test the analysis pipeline. Our mock fields were created using the theoretical angular power spectra as inputs for the \texttt{healpy.synalm} function. The procedure for simulating $N$ correlated fields is described in \cite{Giannantonio_2008} and internally implemented in the \texttt{healpy.synalm} function. The $a_{\ell m}$ were then converted to maps with the \texttt{healpy.alm2map} function.

In our case, we considered $N = 5$ fields: the density field and the ARFs for the two Quaia redshift bins, as well as CMB lensing. These are generated from the $N (N+1)/2 = 15$ input auto- and cross-correlations between the five fields. To generate realistic covariance matrices, we implemented a correction in the autocorrelation power spectra $C_\ell^{gg}$ and $C_\ell^{zz}$ to account for the excess of measured power at the lower multipoles in the real data compared to the theory model due to potential systematics. This correction was applied only to generate mocks that reproduce the real data measurements and derive the covariance matrix. The implementation was done using the following power-law parameterization of the measured power spectra:
    \begin{equation}
    \label{eq:sys}
        C_\ell^{XX,{\rm measured}} \simeq C_\ell^{XX,{\rm theory}} \left[ 1+ \left(\frac{\ell_0}{\ell} \right)^\beta \right],
    \end{equation}
    where $\ell_0$ and $\beta$ are free parameters determined from the data. By fitting with a least squares method the data to Eq.~\ref{eq:sys}, we obtained the $\ell_0$ and $\beta$ parameters listed in Table~\ref{tab:fit} for each of the redshift bins and fiducial bias models. We compare in Fig.~\ref{fig:auto} the data and the theoretical model after applying this correction.

\begin{table}[]
    \centering
    \caption{Derived $\ell_0$ and $\beta$ parameters.}
    \begin{tabular}{|c|c|c|c|c|c|}
    \hline
       \multicolumn{2}{|c|}{Observable} & \multicolumn{2}{|c|}{$C_\ell^{gg}$} & \multicolumn{2}{|c|}{$C_\ell^{zz}$} \\ \hline
       \multicolumn{2}{|c|}{Bin}  & Low-$z$ & High-$z$ & Low-$z$ & High-$z$ \\ \hline 
\multirow{2}{*}{P24 $b(z)$} & $\ell_0$ & 17 & 16 & 5 & 20 \\ 
& $\beta$ & 1.73 & 1.19 & 2.99 & 3.59  \\ \hline
\multirow{2}{*}{L17 $b(z)$} & $\ell_0$ & 32 & 27 & 10 & 20 \\ 
& $\beta$ & 1.50 & 1.87 & 1.87 & 3.57  \\ \hline
    \end{tabular}
    \tablefoot{Best-fit $\ell_0$ and $\beta$ parameters of the model adopted for the excess of observed power in the $C_\ell^{gg}$ and $C_\ell^{zz}$ data for the P24 and L17 fiducial bias evolution models.}
    \label{tab:fit}
\end{table}

After generating the mock signal for the $N=5$ correlated fields, we simulated the noise component. We included in the CMB lensing field another Gaussian map component based on the $Planck$ PR4 lensing theoretical noise. We verified that this approach was compatible at the $\lesssim 10\%$ level with the real measured $C_\ell^{\kappa \kappa}$ from the Planck PR4 data. For the density field, we converted the overdensity to the predicted number counts per pixel using the Quaia data average density $\bar{N}$ and generated a discrete number of QSOs in each pixel by Poisson sampling. For the computation of the ARF maps, we assigned a redshift to each QSO by random sampling a Gaussian distribution centered in $\bar{z}+\delta z$ with a dispersion equal to the RMS of the redshift bins in the Quaia catalog, where $\delta z$ is the predicted ARF field in the pixel.

We finally obtained the joint covariance matrix for our analysis by computing the power spectra of the 1,000 correlated simulations. We also cross-checked that the shot noise estimations with random values for $C_\ell^{gg}$ and $C_\ell^{zz}$ were accurate by recovering the input theoretical power spectra after subtracting the noise within $\lesssim 0.1\sigma$ relative to the standard deviation on the measured $C_\ell^{gg}$ and $C_\ell^{zz}$ from the mocks.
\begin{figure*}
\centering
\includegraphics[width=\textwidth]{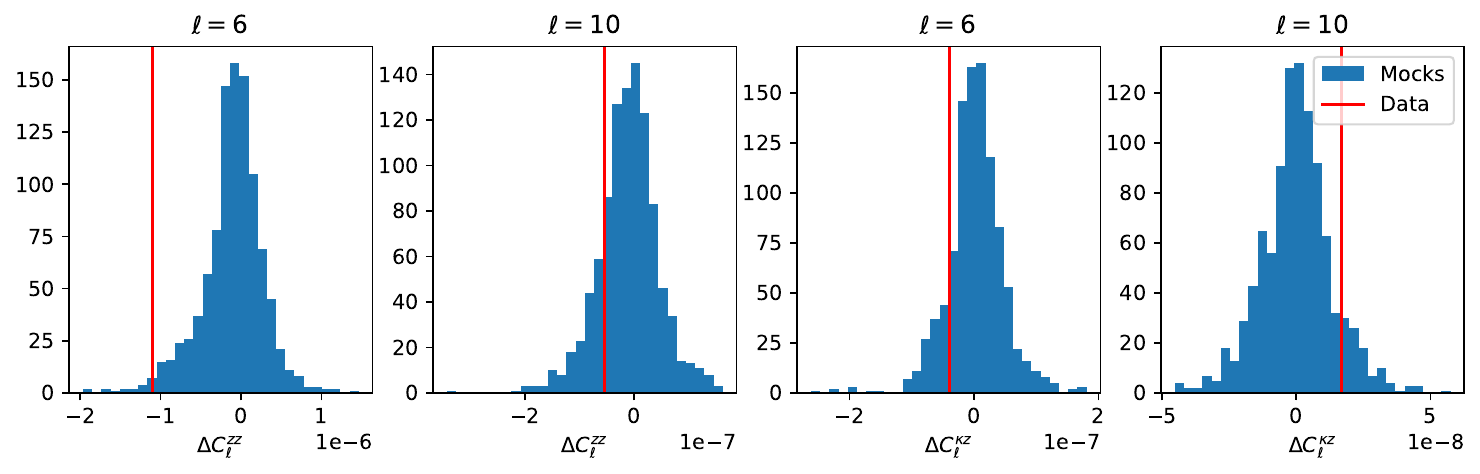}
    \caption{Comparison of the shifts in the $\ell = 6$ and $\ell = 10$ angular power spectra of the Quaia high-z sample after applying systematics deprojection to the mocks and data for $C_\ell^{zz}$ (left panels) and $C_\ell^{\kappa z}$ (right panels), used to validate the scale cuts in our analysis. The blue bars represent the distribution of the shift in the 1000 mock realizations, while the red lines indicate the shift found in real data.}
    \label{fig:sys}
\end{figure*}

\subsection{Systematics deprojection and scale cuts}
\label{sec:sys}
We performed systematics deprojection tests on the data and on the mocks to validate the lower minimum multipoles that can be safely included in our analysis for each observable. For this, we used a joint linear regression for the full set of Quaia systematics templates based on the ordinary least squared (OLS) method introduced in, for example, \cite{weayerdyck}. We applied the linear deprojection to our data and to our 1000 mocks, which do not contain any systematics. Thus, the variations in the angular power spectra, after deprojecting systematics from the mocks, can be attributed to random fluctuations. In this way, we compared the $\Delta C_\ell$ difference at each scale for the data to the $\Delta C_\ell$ distribution for the mocks and determine that a given multipole is affected by systematics if its variation cannot be explained as statistical fluctuation at the 95\% confidence level.

\citet{fabbian2025} note that the measured $C_\ell^{\kappa g}$ from Quaia and $Planck$ data present compatible deviations from the model with random fluctuations at every multipole. Moreover, $C_\ell^{gg}$ present offsets at large scales ($\ell \lesssim 15$) that cannot be explained in terms of statistical fluctuation. Here, for the density and CMB lensing cross-correlation, we adopted the same validated scale cuts in \citet{fabbian2025} and set $\ell_{\rm min} = 14$ for $C_\ell^{gg}$ and $\ell_{\rm min}$ = 2 for $C_\ell^{\kappa g}$. For the ARFs, we validated the cuts with our regression method and find that for $C_\ell^{\kappa z}$ the offset in the data is compatible with statistical fluctuations at every multipole. By contrast, for $C_\ell^{zz}$, scales lower than $\ell \lesssim 10$ present greater offsets than those compatible with the mocks at the 95\% confidence level. In Fig.~\ref{fig:sys} we show an example of this validation for the shift in $C_\ell^{zz}$ and $C_\ell^{\kappa z}$ at $\ell = 6$ and $\ell = 10$, using the Quaia high-z sample.

Angular redshift fluctuations could have a different sensitivity to systematics, since they rely not only on the number of QSOs detected in a given pixel but also on their measured photo-$z$s. The quality of the photo-$z$ estimates typically depends upon photometric conditions that also modulate the number of detected sources, such as air mass, sky background, and extinction. Nevertheless, other aspects may exist, which could modulate the photo-$z$ quality. In the particular case of QSOs, the real QSO redshift represents one aspect: line confusion -- the main cause for QSO catastrophic photo-$z$ assignments -- is more likely for QSOs below $z\approx 2.1$ than for QSOs with this redshift, since for the latter the (bright) Lyman-$\alpha$ enters Gaia's optical system. Such modulation should be present mainly in the lower redshift bin under consideration, which has a lower weight given the redshift-dependent bias in our analysis.

Note that our baseline angular power spectra for the analysis are those prior to systematics deprojection. Since the measured angular power spectra after deprojection for multipoles larger than our scale cuts is compatible with random fluctuations, we deemed the removal of the lower multipoles sufficient to mitigate the observational systematics.

\subsection{Likelihood and parameter inference}

We define the likelihood ${\cal L}$ as a multivariate Gaussian
\begin{equation}
   -2 \log {\cal L} \equiv \chi^2 = \sum_{\ell,\ell'} \left(C_\ell^{{\rm obs}} - \Tilde{C}_\ell (\theta) \right){\rm Cov}^{-1}_{\ell \ell'}\left(C_{\ell'}^{{\rm obs}} - \Tilde{C}_{\ell'} (\theta) \right),
\end{equation}
where $C_\ell^{{\rm obs}}$ are the elements of the data vector, $\Tilde{C}_\ell (\theta)$ is the theoretical model of the angular power spectrum for a given set of parameters $\theta$, and ${\rm Cov}^{-1}$ is the inverse of the covariance matrix. We consider the Gaussian likelihood approximation safe enough, since \cite{Alonso_2023} demonstrate that the distribution of the QSO - CMB lensing cross-correlation measurements (and the autocorrelation for small scales) are well described by a Gaussian. Our theoretical model is based on the \texttt{ARFCAMB} code described in Sect.~\ref{sec:model}. To account for possible uncertainties on the galaxy bias modeling, we adopted an extension of the parameterization of the Quaia $b(z)$ by \citet{Piccirilli_2024} expressed in Eq.~\ref{eq:bzquaia}:
\begin{equation}
    b(z) = \frac{b_0}{D(z)} (1+z)^\alpha,
\end{equation}

where $b_0$ and $\alpha$ are free parameters and the $b_0 = 1.26, \alpha = 0$ case reduces to the fiducial bias evolution model by \cite{Piccirilli_2024}. To compute the constraints on the cosmological parameters, we implemented our likelihood using the MCMC sampler \texttt{emcee}\footnote{\url{https://emcee.readthedocs.io/}} \citep{Foreman_Mackey_2013}. Our analysis includes $f_{\rm NL}$ as well as the bias parameters $b_0$ and $\alpha$ as the three cosmological parameters to constrain. The rest of the cosmological parameters were fixed to the $Planck$ 2018 cosmology \cite{1807.06209}.

We used as minimum multipoles $\ell_{\rm min} = 2$ for $C_\ell^{\kappa g}$ and $C_\ell^{\kappa z}$ and $\ell_{\rm min} = 14$ for $C_\ell^{gg}$ and $\ell_{\rm min} = 10$ for $C_\ell^{zz}$, according to the tests discussed in Sect.~\ref{sec:sys}. For the maximum multipoles, we set $\ell_{\rm max} = 300$ for the high-z redshift bin, and we cut the Low-z bin to $\ell_{\rm max} = 240$ to avoid including $k$-modes greater than $k\sim 0.07~h$~Mpc$^{-1}$. Although for the high-z bin the $k \sim 0.07~h$~Mpc$^{-1}$ cutoff would correspond to $\ell \lesssim  400$, we decided not to include any multipole larger than 300 in the analysis given the limited resolution of our maps ($N_{\rm side} = 128$). In our theoretical model from \texttt{ARFCAMB}, we also included nonlinear contributions to the power spectrum using \texttt{halofit} \citep{Takahashi_2012}. Furthermore, to make proper comparisons with the theory model, we corrected our data for the pixel window function, given that its impact is nonnegligible for $\ell_{\max} \gtrsim 2 N_{\rm side}$.

\section{Results}
\label{sec:results}

We calculated the constraints on $f_{\rm NL}$ and the $b_0$, $\alpha$ bias model parameters individually for the four observables involved in our analysis ($C_\ell^{gg}$, $C_\ell^{\kappa g}$, $C_\ell^{zz}$, and $C_\ell^{\kappa z}$). The median likelihood values at the 68\% confidence intervals are reported in Table~\ref{tab:results}, together with the two combinations of density and ARFs with their respective CMB lensing cross-correlations and joint analysis of the four observables. The measured constraints before adding ARFs, only considering $C_\ell^{gg} + C_\ell^{\kappa g}$, are consistent with the findings by \cite{fabbian2025},  who report $f_{\rm NL} = -20.5^{+19.0}_{-18.1}$ assuming the universality relation with $p=1$. As in \cite{fabbian2025}, our joint constraint from density and CMB lensing is dominated by $C_\ell^{\kappa g}$ due to the Poisson shot noise limitations of the Quaia sample for $C_\ell^{gg}$. By contrast, for the $C_\ell^{zz} + C_\ell^{\kappa z}$ constraints, we find that the ARF autocorrelation $C_\ell^{zz}$ is the dominant term in the joint analysis. This makes sense provided that, for ARFs, the cross-angular power spectrum multipoles $C_\ell^{\kappa z}$ are very close to zero for all but the low $\ell$s, which are dominated by cosmic variance. We measure $f_{\rm NL} = -43^{+50}_{-44}$ from ARFs and their cross-correlation with CMB lensing. Moreover, when we combine this information with Quaia density and its cross-correlation with CMB lensing, we find $f_{\rm NL} = -3 \pm 14$. The result represents a $\sim$25\% improvement with respect to the measurement in \citet{fabbian2025} for $C_\ell^{gg} +C_\ell^{\kappa g}
$. This is consistent with the expected theoretical impact of ARFs in the constraints (see Appendix~\ref{sec:mockdata} for more details). Our measurement is the second most precise constraint on $f_{\rm NL}$ from LSS using two-point functions to date, after the \citet{Chaussidon_2025} measurement using DESI DR1 spectroscopic data ($\sigma(f_{\rm NL}) \sim 9$). 

\begin{figure}
\centering
\includegraphics[width=\columnwidth]{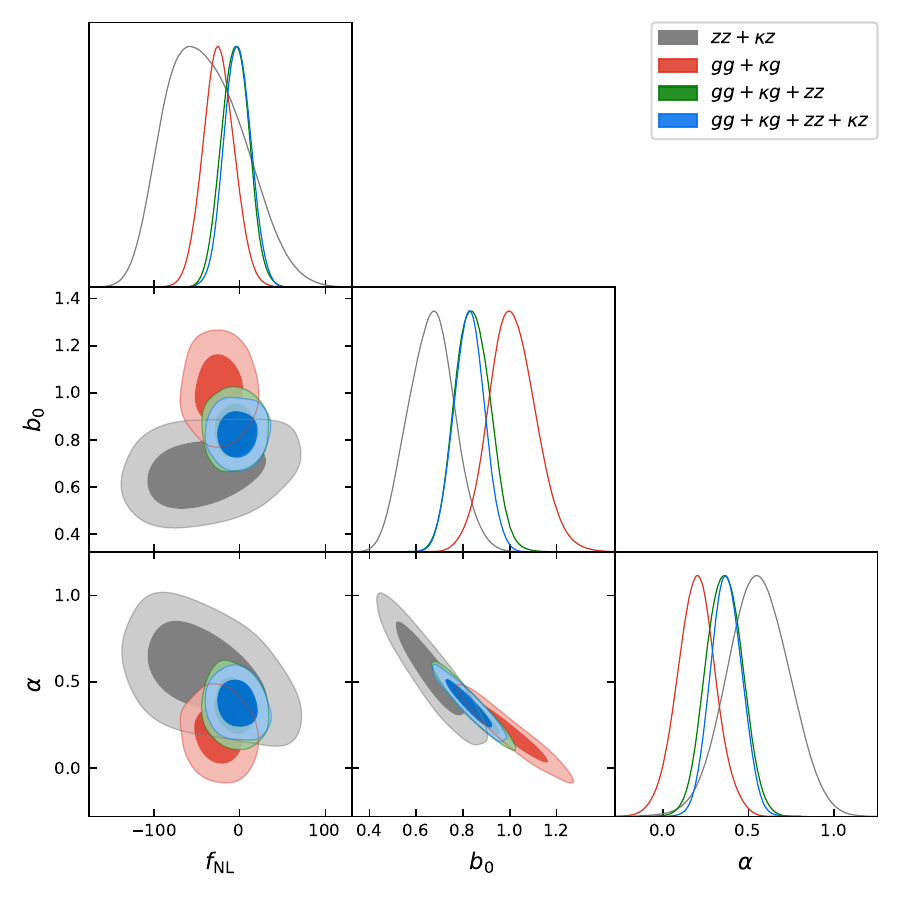}
    \caption{1$\sigma$ and 2$\sigma$ confidence ellipses for $f_{\rm NL}$ and the $b_0$ and $\alpha$ bias parameters measured from Quaia and $Planck$ data, assuming the same bias parameterization for density and ARFs (baseline case). The gray contours represent the constraints from ARFs plus their CMB lensing cross-correlation, the red contours represent the density plus their CMB lensing cross-correlation, the green contours represent the combination of density, ARF, and density - CMB lensing cross-correlation, and the blue contours represent the combination of density, ARFs, and their CMB lensing cross-correlations.}
    \label{fig:fnl}
\end{figure}

In Table~\ref{tab:results} the $\chi^2$ per degrees of freedom (dof) estimates in the right column indicate mild tensions between the best-fit model and the observed data, particularly for those combinations including $C_\ell^{\kappa z}$. For this observable, this is mainly due to a few data points (e.g., the $\ell\simeq 12$ data point, which lies more than 3$\sigma$ from the preferred models; see Fig.~\ref{fig:cross}). We verified that if we removed these data points from the analysis, the reduced $\chi^2$ values of the combined analysis would improve (see more details in Appendix \ref{sec:chi2}). If we did not include any information from $C_\ell^{\kappa z}$, which presents a worse $\chi^2$/dof than the rest of the observables, the constraint from adding $C_\ell^{zz}$ only to the density and CMB lensing information would be $f_{\rm NL} = -5^{+16}_{-15}$. We represent in Fig.~\ref{fig:fnl} the comparison of the constraints on the parameters from $C_\ell^{gg}+C_\ell^{\kappa g}$, $C_\ell^{zz}+C_\ell^{\kappa z}$ and their combinations, including the case where $C_\ell^{\kappa z}$ is neglected. We observe moderate tension ($\lesssim 2~\sigma$) between the bias parameter best-fit estimates from $C^{zz}_\ell$ and $C^{\kappa z}_\ell$. The tension also appears when we compare the combinations $C_{\ell}^{gg} + C_\ell^{\kappa g}$ and  $C_{\ell}^{zz} + C_\ell^{\kappa z}$. We believe this is caused by the higher sensitivity of the ARF kernel to higher $k$-modes (or nonlinear scales), which impacts $C^{zz}_\ell$ more than $C^{\kappa z}_\ell$. We discuss this further below.

\begin{table}[]
    \centering
    \caption{Constraints on cosmological parameters.}
    \begin{tabular}{|c|c|c|c|c|}
    \hline
Observ. & $f_{\rm NL}$ & $b_0$ & $\alpha$ & $\chi^2/$dof\\ \hline
$C_\ell^{gg}$ & -51$^{+37}_{-39}$ & 0.93 $\pm$ 0.14& 0.33$^{+0.18}_{-0.16}$ & 42/45\\ \hline 
$C_\ell^{zz}$ & -93$^{+43}_{-26}$ & 0.53$^{+0.13}_{-0.12}$ & 0.83$^{+0.25}_{-0.26}$ & 39/49 \\ \hline 
$C_\ell^{\kappa g}$ & -16$^{+22}_{-23}$ & 0.96$^{+0.15}_{-0.13}$ & 0.23 $\pm$ 0.17 & 63/57 \\ \hline 
$C_\ell^{\kappa z}$ & 187$^{+516}_{-144}$  &  1.49$^{+0.77}_{-0.54}$ & 0.02$^{+0.29}_{-0.45}$ & 105/57 \\ \hline 
$C_\ell^{gg}$+$C_\ell^{\kappa g}$ & -25$^{+19}_{-17}$ & 1.01$^{+0.10}_{-0.09}$  & 0.20$_{-0.10}^{+0.11}$ & 113/105 \\ \hline 
$C_\ell^{zz}$+$C_\ell^{\kappa z}$ & -43$^{+50}_{-44}$ & 0.67$^{+0.09}_{-0.10}$ & 0.55$^{+0.19}_{-0.17}$  & 164/109 \\ \hline 
$C_\ell^{gg}$+$C_\ell^{\kappa g}$ & \multirow{2}{*}{-5$^{+16}_{-15}$} & \multirow{2}{*}{0.82 $\pm$ 0.07} &  \multirow{2}{*}{0.39 $\pm$ 0.10} & \multirow{2}{*}{176/157} \\  
 + $C_\ell^{zz}$ & & & & \\ \hline
$C_\ell^{gg}$+$C_\ell^{\kappa g}$ & \multirow{2}{*}{-3 $\pm$ 14} & \multirow{2}{*}{0.83 $\pm$ 0.06} &  \multirow{2}{*}{0.37 $\pm$ 0.09} & \multirow{2}{*}{338/217} \\  
 + $C_\ell^{zz}$+$C_\ell^{\kappa z}$ & & & & \\ \hline

\end{tabular}
    \tablefoot{Median likelihood cosmological parameters with 1$\sigma$ confidence intervals for the baseline analysis from the four observables involved in our analysis and their combinations. We also list the $\chi^2$ per dof for each case.}
    \label{tab:results}
\end{table}

To investigate whether these tensions could be related to degeneracies between parameters and to test the stability of the $b_0$ and $\alpha$ measurements, we recomputed the constraints on the three parameters from $C_\ell^{gg}+C_\ell^{\kappa g}$ and $C_\ell^{zz}+C_\ell^{\kappa z}$ using $\ell_{\rm min} = 20$ and $\ell_{\min} = 40$. The purpose of these scale cuts is to remove the larger scales where the $f_{\rm NL}$ sensitivity arises, leaving only small-scale information which is sensitive to the galaxy bias. We list the constraints with these $\ell_{\rm min}$ cuts and compare them to the baseline case with lower $\ell_{\rm min}$ in Table~\ref{tab:lmin}. We find the constraints from $C_\ell^{zz}+C_\ell^{\kappa z}$ are stable against the scale cuts. For $C_\ell^{gg}+C_\ell^{\kappa g}$, we find hints of a trend in which we measure larger $\alpha$ parameters with the higher $\ell_{\rm min}$ cuts. Although it is not statistically significant, this trend is consistent with that expected when more weight is given to smaller, potentially nonlinear scales.

\begin{figure}
\centering
\includegraphics[width=\columnwidth]{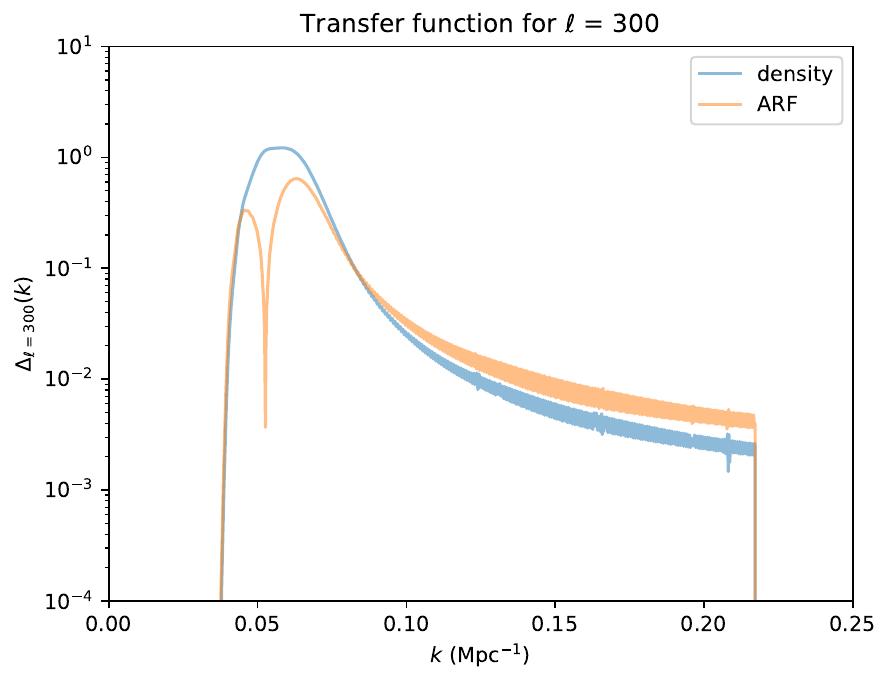}
    \caption{Transfer function for the density (blue line) and ARF (orange line) fields obtained assuming the high-z Quaia sample for $\ell = 300$, as a function of the $k$ scale.}
    \label{fig:transfer}
\end{figure}

To understand whether the tension on the bias parameters could be related to limits on the modeling of nonlinear scales, we also computed the constraints from  $C_\ell^{gg}+C_\ell^{\kappa g}$ and $C_\ell^{zz}+C_\ell^{\kappa z}$ with an $\ell_{\rm max} = 200$ cut. We compare in Table~\ref{tab:lmax} the results for the baseline case. We find that the measurement from $C_\ell^{gg}+C_\ell^{\kappa g}$ is stable with $\ell_{\rm max}$; by contrast, the bias parameters measured from $C_\ell^{zz}+C_\ell^{\kappa z}$ vary when adopting a lower $\ell_{\rm max}$ and the tension between observables is alleviated. This suggests that the discrepancy on the bias measurement is related to the higher impact of smaller (and potentially nonlinear) scales on the ARF kernel under the same redshift window function. This is somewhat expected, since the ARF kernel can be seen as a radial gradient of the 3D density field under the redshift shell, and, unlike the density kernel, is sensitive to variations of the QSO density field under the redshift window \citep[see][for further details]{Hern_ndez_Monteagudo_2020}. To illustrate this, in Fig.~\ref{fig:transfer} we represent the transfer function for $\ell = 300$ assuming the high-$z$ Quaia sample, where we show that for ARFs the sensitivity of the transfer function for higher $k$-modes is greater compared to the density. This is due not only due to its higher amplitude at high $k$ values but also due to the sign flip at $k\sim 0.07~$Mpc$^{-1}$. This suggests that future work will be needed to improve the modeling of nonlinear scales in ARFs; however, our scope in this work is to effectively quantify the impact on the $f_{\rm NL}$ determination of these sources of systematics.
 
\begin{table}[]
    \centering
    \caption{Constraints on parameters as a function of $\ell_{\rm min}$.}
    \begin{tabular}{|c|c|c|c|c|}
    \hline
Observables & $\ell_{\rm min}$ & $f_{\rm NL}$ & $b_0$ & $\alpha$ \\ \hline

\multirow{3}{*}{$C_\ell^{gg}$+$C_\ell^{\kappa g}$} 
& 14, 2 & -25$^{+19}_{-17}$ & 1.01$^{+0.10}_{-0.09}$  & 0.20$_{-0.10}^{+0.11}$ \\ 
&  20 & -93$^{+29}_{-25}$ & 0.83 $\pm$ 0.07  & 0.45 $\pm$ 0.11 \\ 
&  40 & -43$^{+44}_{-45}$ & 0.77 $\pm$ 0.07  & 0.52$^{+0.10}_{-0.12}$\\ \hline

\multirow{3}{*}{$C_\ell^{zz}$+$C_\ell^{\kappa z}$}
& 10, 2 & -43$^{+50}_{-44}$ & 0.67$^{+0.09}_{-0.10}$ & 0.55$^{+0.19}_{-0.17}$\\
& 20 & -16$^{+59}_{-56}$ & 0.75$^{+0.08}_{-0.07}$ & 0.42$^{+0.14}_{-0.17}$\\
& 40 & -36$^{+86}_{-76}$ & 0.79$^{+0.12}_{-0.13}$ & 0.37$^{+0.20}_{-0.19}$ \\ \hline

\end{tabular}
    \tablefoot{Median likelihood cosmological parameters with 1$\sigma$ confidence intervals for the $C_\ell^{gg}+C_\ell^{\kappa g}$ and $C_\ell^{zz}+C_\ell^{\kappa z}$ observables as a function of the minimum multipole(s) used in the analysis.}
    \label{tab:lmin}
\end{table}

\begin{table}[]
    \centering
    \caption{Constraints on parameters as a function of $\ell_{\rm max}$}
    \begin{tabular}{|c|c|c|c|c|}
    \hline
Observables & $\ell_{\rm max}$ & $f_{\rm NL}$ & $b_0$ & $\alpha$ \\ \hline

\multirow{2}{*}{$C_\ell^{gg}$+$C_\ell^{\kappa g}$} 
& 240, 300 & -25$^{+19}_{-17}$ & 1.01$^{+0.10}_{-0.09}$  & 0.20$_{-0.10}^{+0.11}$\\ 
& 200 & -9$^{+29}_{-26}$ & 1.03$^{+0.13}_{-0.12}$  & 0.12 $\pm$ 0.15\\ \hline 
\multirow{2}{*}{$C_\ell^{zz}$+$C_\ell^{\kappa z}$}
& 240, 300 & -49$^{+53}_{-36}$ & 0.63$^{+0.10}_{-0.08}$ & 0.60 $\pm$ 0.17 \\
& 200 & 95$^{+117}_{-89}$ & 0.95$^{+0.16}_{-0.13}$ & 0.02$^{+0.22}_{-0.25}$ \\ \hline

\end{tabular}
    \tablefoot{Median likelihood cosmological parameters with 1$\sigma$ confidence intervals for the $C_\ell^{gg}+C_\ell^{\kappa g}$ and $C_\ell^{zz}+C_\ell^{\kappa z}$ observables as a function of the maximum multipole used in the analysis.}
    \label{tab:lmax}
\end{table}

\begin{figure*}
\centering
\includegraphics[width=0.8\textwidth]{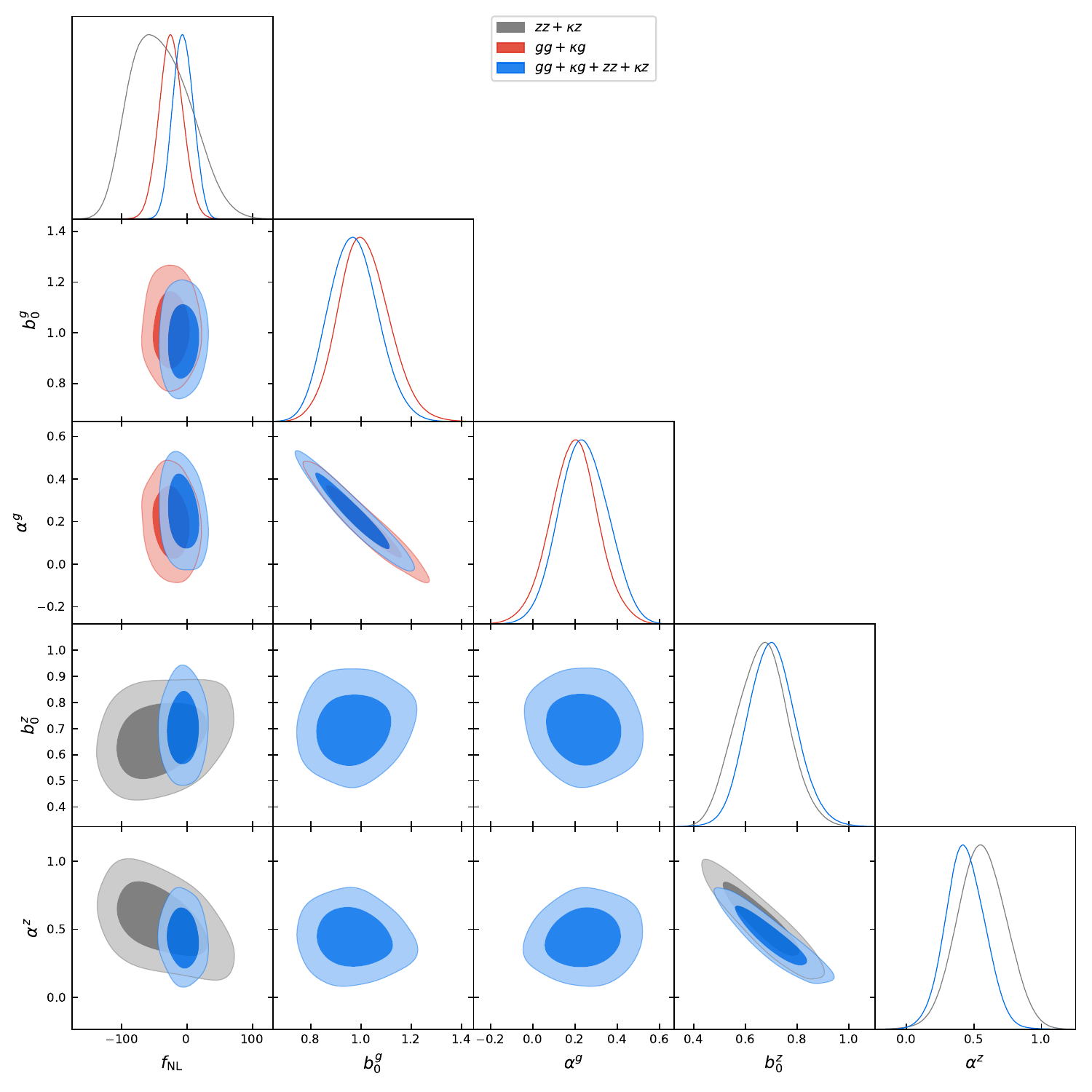}
    \caption{1$\sigma$ and 2$\sigma$ confidence ellipses for $f_{\rm NL}$ and the $b_0$ and $\alpha$ bias parameters measured from Quaia and $Planck$ data, assuming two independent bias parameterizations for density and ARFs. The gray contours represent the constraints from ARFs plus their CMB lensing cross-correlation, the red contours represent the density plus their CMB lensing cross-correlation, and the blue contours represent the combination of density, ARFs, and their CMB lensing cross-correlations.}
    \label{fig:fnl2b}
\end{figure*}

To quantify the effect on the constraints on $f_{\rm NL}$ of the assumptions related to the fiducial galaxy bias evolution, we recomputed the joint constraints on $f_{\rm NL}$ including all observables changing the baseline assumptions. The tests included the assumption of an underlying bias redshift evolution following $\alpha = 0$, as measured in P24, and the use of a different bias evolution such as the eBOSS model by L17. Furthermore, to test whether the tensions on $b_0$ and $\alpha$ between observables caused by nonlinear scales modeling impact the final results, we also considered a case in which we varied two pairs of the $b_0$ and $\alpha$ parameters, one for $C_\ell^{gg}$ and $C_\ell^{\kappa g}$ and another for $C_\ell^{zz}$ and $C_\ell^{\kappa z}$, in order that the nonlinear physics would be parameterized in an effective bias for each field. The results of these tests are listed in Table~\ref{tab:testbz}. For the P24 and L17 bias models, rather than varying $b_0$ and $\alpha$, we fit two amplitude parameters $A_1$ and $A_2$ for each redshift bin. For all the tests, we obtain compatible results with almost no shift in the measured $f_{\rm NL}$ and similar confidence intervals. We conclude that our result is robust against the fiducial assumptions on the $b(z)$ evolution model. In particular, for the test with two pairs of bias parameters, we find $f_{\rm NL} = -6_{-14}^{+15}$, a similar error bar to the baseline case ($f_{\rm NL} = -3 \pm 14)$. The values of each $b_0$ and $\alpha$ are compatible with those of the baseline cases for $C_\ell^{\rm gg} + C_\ell^{\kappa g}$ and $C_\ell^{\rm zz} + C_\ell^{\kappa z}$ reported in Table~\ref{tab:results}. This shows that our constraints on $f_{\rm NL}$ are stable when the shift in the $b_0$ and $\alpha$ parameters is reduced or eliminated. Thus, our results are not artificially enhanced by the tension between observables on the bias parameters. We show the constraints on the parameters assuming two separate bias parameterizations in Fig.~\ref{fig:fnl2b}.

We assumed the universality relation with $p=1$ as a baseline for computing the theoretical $f_{\rm NL}$ impact on the scale-dependent bias. However, several studies in the literature (e.g., \citealt{Slosar_2008}) state that $p=1.6$ is a more realistic value for the response of the QSO bias to $f_{\rm NL}$. Furthermore, recent studies such as \citet{Barreira_2020,Barreira_2022} state that given the uncertainties in the $p$ parameter we can only determine $b_\phi f_{\rm NL}$ through LSS measurements. To assess the impact on the $f_{\rm NL}$ constraints of a potentially lower QSO response to the local PNG parameter, we recomputed the constraints from the joint analysis of all observables assuming $p=1.6$. We also measured the independent $b_\phi f_{\rm NL}$ constraints from the same data. We obtain $f_{\rm NL} = 2^{+20}_{-21}$ for the $p=1.6$ case and $b_\phi f_{\rm NL} = -63^{+79}_{-89}$ for the model-independent assumption.

In summary, we show that the improvement in the $f_{\rm NL}$ uncertainty due to the addition of ARFs as a new observable is robust enough against the possible systematics in our analysis. The internal tensions between the observables suggest that ARFs are more sensitive to nonlinear physics under the same $\ell_{\rm max}$ scale cuts than 2D clustering, and future work will be needed to better model the nonlinearities for ARFs. However, in this work, we have shown that when parameterizing these uncertainties with an independent effective bias, our $f_{\rm NL}$ measurement remains unaffected. Our results are promising as the first application of this new cosmological observable to measuring $f_{\rm NL}$ from real data. They indicate that ARFs should be included in the future analysis pipelines of 2D clustering, in particular for surveys that will be analyzed in 2D harmonic space such as LSST or the Euclid photometric survey. On the other hand, a 2D analysis of spectroscopic surveys such as DESI could be improved with the inclusion of ARFs, which should be a more robust observable with zero or negligible photometric redshift uncertainties. This will provide a 2D alternative to the standard 3D clustering analysis with additional constraining power.

\begin{table}[]
    \caption{Constraints on parameters for the various bias assumptions.}
    \centering
    \begin{tabular}{|c|c|c|c|}
    \hline
         & $f_{\rm NL}$ & $b_0$  & $\alpha$ \\ \hline
      Baseline &  -3 $\pm$ 14 & 0.83 $\pm$ 0.06   & 0.37 $\pm$ 0.08 \\ \hline
        & $f_{\rm NL}$ & $b_0^g$ $(b_0^z)$  & $\alpha^g$ $(\alpha^z)$ \\ \hline
     \multirow{2}{*}{Separate $b_0$ and $\alpha$} & \multirow{2}{*}{-6$^{+15}_{-14}$}  & 0.92 $\pm$ 0.09 & 0.27 $\pm$ 0.12  \\
       & & (0.73 $\pm$ 0.10)   & (0.41 $\pm$ 0.14) \\ \hline
      & $f_{\rm NL}$ & $A_1$ & $A_2$ \\ \hline
        P24 $b(z)$ & -2$^{+14}_{-16}$ & 1.05 $\pm$ 0.03 & 1.24 $\pm$ 0.04 \\
        L17 $b(z)$ & -4 $\pm $ 14 & 1.05 $\pm$ 0.03& 0.97$^{+0.03}_{-0.04}$ \\ \hline  
    \end{tabular}
    \tablefoot{Median likelihood cosmological parameters with 1$\sigma$ confidence intervals for the baseline analysis of the combination of observables compared to different assumptions on the $b(z)$ fiducial evolution and number of parameters.}
    \label{tab:testbz}
\end{table}

\section{Conclusions}
\label{sec:conclusions}
In this work we studied for the first time on real data
the sensitivity of ARFs -- a new 2D
observable that measures matter tracers' redshift angular deviations as a cosmological probe --
to the PNG parameter $f_{\rm NL}$. Using the Quaia QSO catalog, constructed from Gaia and unWISE data, we extended a previous analysis in which $f_{\rm NL}$ was measured from Quaia QSO density cross-correlated with the $Planck$ PR4 CMB lensing to include ARFs as an additional cosmological probe.

We developed an analysis pipeline based on a modification of the \texttt{CAMB} Boltzmann code, and we used that theoretical model to generate correlated simulations of the three fields involved in our analysis: CMB lensing, density, and ARFs. To deal with the possible systematics in the auto-power spectra of density and ARFs, we implemented scale cuts in the analysis and fit a power-law model to the measured auto power spectra to generate a realistic covariance matrix.

Assuming the universality relation with $p=1$, solely from Quaia ARFs cross-correlated with the $Planck$ CMB lensing, we obtain $f_{\rm NL} = -43^{+50}_{-44}$. We also measure $f_{\rm NL} = -25^{+19}_{-17}$ from the Quaia density cross-correlated with $Planck$ lensing, a constraint that is fully consistent with the results in \cite{fabbian2025}. By combining Quaia density, ARFs, and their respective CMB lensing cross-correlations, we obtain as a joint constraint $f_{\rm NL} = -3 \pm 14$, which represents a $\sim$ 25\% improvement thanks to the addition of ARFs. If we do not include the ARF - CMB lensing cross-correlation for the sake of a better goodness-of-fit, we measure $f_{\rm NL} = -5^{+16}_{-15}$, yet again improving the constraints from Quaia density and CMB lensing by $\sim 17\%$ . Despite the internal tensions on the measured galaxy bias parameters from different observables, we verified that our constraint is robust enough against the different assumptions for the evolution of the galaxy bias, including fitting an independent effective bias model for ARFs. Our result is the second tightest constraint on $f_{\rm NL}$ using two-point LSS functions to date and the most precise obtained from projected 2D summary statistics. 

Our study supports the inclusion of ARFs as an extra cosmological probe in 2D analysis in the harmonic space with data from ongoing and future large-scale surveys such as DESI, Euclid, and LSST. Among the future work needed to improve analyses involving ARFs, we note the need for improved understanding of ARF systematics, which may be particularly important when using photometric redshifts. This should allow us to include observables such as the cross-correlation between density and ARFs. Furthermore, improved modeling of nonlinearities in the ARFs will be key to performing a joint analysis with 2D galaxy densities, given the higher sensibility of the ARF kernel to nonlinear physics under the same multipoles.

\begin{acknowledgements}
JRBC, CHM, ACP, and JMC acknowledge the support of the Spanish Ministry of Science and Innovation under the grants PID2021-126616NB-I00 and ``DarkMaps'' PID2022-142142NB-I00, and from the European Union through the grant ``UNDARK'' of the Widening participation and spreading excellence program (project number 101159929). JRBC acknowledges support from a Momentum MSCA Fellowship, co-funded by the European Comission through the HORIZON-MSCA-2023-COFUND programme and the Secretariat of the Hungarian Academy of Sciences (MTA). We thank the members of the Quaia team and F. Bouchet for helpful feedback and discussions.
\end{acknowledgements}

\bibliographystyle{aa}
\bibliography{bib.bib}

\begin{appendix}
\section{Investigating the reduced $\chi^2$ behavior}
\label{sec:chi2}
The constraints reported in Table~\ref{tab:results} show that the $\chi^2$/dof values of the fit for the combined analysis of the four observables would result in a very low probability-to-exceed (PTE) value. We investigate here which data points drive this behavior and estimate again the constraints obtained by removing these data points to test the robustness of our joint measurement. In particular, we filter out the data points that are more than 2.5$\sigma$ off with respect to the best-fit model. These are in total 6 data data points, corresponding to the multipole bins centered at $\ell = 194.5$ for the $C_\ell^{\kappa g}$ ``Low-z" sample,  $\ell = 30.5$ for the $C_\ell^{\kappa g}$ ``High-z" sample and $\ell = 12.5, 47.0, 194.5, 224.5$ for the $C_\ell^{\kappa z}$ ``Low-z" sample. 

As already mentioned in Sect.~\ref{sec:results}, most of the tension with the best-fit model is coming from $C_\ell^{\kappa z}$ data points; for instance, the $\chi^2/$dof for this observable alone is 105/57 (see Table~\ref{tab:results}) and this is propagated to every combination including $C_\ell^{\kappa z}$. However, the constraining power of the CMB lensing - ARF cross-correlation for our datasets is quite limited and the final impact in the analysis of removing these data points is not significant. We list in Table~\ref{tab:chi2} the comparison between the baseline joint analysis and the case filtering the 6 data points that are $> 2.5 \sigma$ off from the best-fit model. The results show full consistency with the baseline analysis and just a slight enlargement of the error bars. For this case, the $\chi^2/$dof is improved to 236/211, which corresponds to a PTE $\sim$ 0.11, just by removing six data points out of 220. This reflects that our results are robust enough despite the reported $\chi^2$ values in Table~\ref{tab:results} and motivates studies with more datasets to better understand the presence of systematic offsets in the cross-correlations with CMB lensing.

\begin{table}[h]
    \centering
    \caption{Constraints on parameters from Quaia data.}
    \begin{tabular}{|c|c|c|c|c|}
    \hline
Data points & $f_{\rm NL}$ & $b_0$ & $\alpha$ & $\chi^2/$dof\\ \hline

Baseline & -3 $\pm$ 14 & 0.83 $\pm$ 0.06   & 0.37 $\pm$ 0.08 & 338/217 \\ \hline 
Filtered & 3 $\pm$ 15 & 0.84 $\pm$ 0.07   & 0.34 $\pm$ 0.10 & 236/211 \\ \hline

\end{tabular}
    \tablefoot{Median likelihood cosmological parameters with 1$\sigma$ confidence intervals from the joint analysis of the four observables involved in our analysis, for the baseline case and the case removing data points in tension. We also list the $\chi^2$ per dof for each case.}
    \label{tab:chi2}
\end{table}

\section{Constraints from Quaia mock data}
\label{sec:mockdata}
In order to validate the MCMC inference pipeline and to provide a realistic estimation of the expected constraints on $f_{\rm NL}$ from the Quaia data and the improvement achieved with the addition of ARFs, we calculate the constraints on the parameters for the four observables involved and their combinations using the average angular power spectra from our 1000 mock realizations. We list the results in Table~\ref{tab:mocks}. Our mocks have as fiducial parameters $f_{\rm NL} = 0$, $b_0 = 1.26$ and $\alpha = 0$. We recover these values with good agreement, and note that in particular for $C_\ell^{\kappa z}$ alone the parameters are essentially unconstrained, as it happens for the real data. Regarding the final constraints achieved after the addition of ARFs, we obtain a $\sim 30\%$ improvement on the $f_{\rm NL}$ error bar. This reflects that the improvement that we report from the data is expected for a theoretical forecast of an ideal case without systematics, and that the $\sim 25\%$ improvement obtained with real data is not a product of tensions on the bias between observables, as already discussed within the text.

\begin{table}[h]
    \centering
    \caption{Constraints on parameters from mock data.}
    \begin{tabular}{|c|c|c|c|}
    \hline
Observable & $f_{\rm NL}$ & $b_0$ & $\alpha$ \\ \hline
$C_\ell^{gg}$ & -8$^{+38}_{-46}$ & 1.30 $\pm$ 0.16& -0.04$^{+0.16}_{-0.15}$ \\ \hline 
$C_\ell^{zz}$ & -10$^{+44}_{-51}$ & 1.30 $\pm$ 0.16 & -0.03$^{+0.17}_{-0.18}$ \\ \hline 
$C_\ell^{\kappa g}$ & 1 $\pm$ 23 & 1.29$^{+0.20}_{-0.16}$ & -0.03$^{+0.16}_{-0.18}$ \\ \hline 
$C_\ell^{\kappa z}$ & 220$^{+1279}_{-526}$  &  1.30$^{+0.76}_{-0.36}$ & -0.61$^{+0.50}_{-0.27}$  \\ \hline 
$C_\ell^{gg}$+$C_\ell^{\kappa g}$ & -1$ \pm$ 20 & 1.27 $\pm$ 0.10  & -0.01$\pm$ 0.09\\ \hline 
$C_\ell^{zz}$+$C_\ell^{\kappa z}$ & 0$^{+34}_{-40}$ & 1.27$^{+0.11}_{-0.10}$ & -0.02$^{+0.12}_{-0.13}$ \\ \hline
$C_\ell^{gg}$+$C_\ell^{\kappa g}$ & \multirow{2}{*}{0 $\pm$ 14} & \multirow{2}{*}{1.26 $^{+0.09}_{-0.08}$} &  \multirow{2}{*}{-0.01 $\pm$ 0.09}  \\  
 + $C_\ell^{zz}$ & & &  \\ \hline
$C_\ell^{gg}$+$C_\ell^{\kappa g}$ & \multirow{2}{*}{0 $\pm$ 14} & \multirow{2}{*}{1.26 $\pm$ 0.07} &  \multirow{2}{*}{-0.01$^{+0.08}_{-0.07}$}  \\  
 + $C_\ell^{zz}$+$C_\ell^{\kappa z}$ & & &  \\ \hline

\end{tabular}
    \tablefoot{Median likelihood cosmological parameters with 1$\sigma$ confidence intervals for the baseline analysis from the four observables involved in our analysis and their combinations applied to Quaia mock data. The fiducial values of the parameters for the mocks are $f_{\rm NL} = 0$, $b_0 = 1.26$ and $\alpha = 0$.}
    \label{tab:mocks}
\end{table}

\end{appendix}

\end{document}